\documentclass[11pt]{article}

\usepackage[T1]{fontenc}
\usepackage{amsmath}
\usepackage{newtxtext,newtxmath}
\usepackage[letterpaper,top=0.88in,bottom=0.90in,
            left=1in,right=1in,headheight=22pt]{geometry}
\usepackage{booktabs}
\usepackage{graphicx}
\usepackage[table]{xcolor}
\usepackage{colortbl}
\usepackage{array}
\usepackage{makecell}
\usepackage{enumitem}
\usepackage{microtype}
\usepackage{float}
\usepackage{setspace}
\usepackage{abstract}
\usepackage{titlesec}
\usepackage{fancyhdr}
\usepackage{caption}
\usepackage{subcaption}
\usepackage{xurl}
\usepackage{etoolbox}
\usepackage[section]{placeins}
\definecolor{papergray}{HTML}{555555}
\usepackage[authoryear,round]{natbib}
\setcitestyle{aysep={,},yysep={;},notesep={; }}
\usepackage[hidelinks]{hyperref}
\usepackage{tikz}
\usetikzlibrary{arrows.meta,positioning,shapes.geometric,calc,decorations.pathreplacing,fit,backgrounds}

\hypersetup{
  pdftitle={Reading the News Without Hindsight: NoLBERT Topics and the Cross-Section of Stock Returns},
  pdfauthor={Kevin Foley, Jonathan Hartadi, Shivesh Prakash, Charishma Takkallapalli, Swapnil Vatsal},
  pdfsubject={Narrative asset pricing with point-in-time language models},
  pdfkeywords={narrative asset pricing, topic models, NoLBERT, contrastive fine-tuning, return prediction, financial news}
}

\setlist{nosep,leftmargin=1.5em}

\titleformat{\section}[hang]
  {\Large\bfseries\color{black}}{\thesection}{0.7em}{}
\titleformat{\subsection}[hang]
  {\large\bfseries\color{black}}{\thesubsection}{0.65em}{}
\titleformat{\subsubsection}[hang]
  {\normalsize\bfseries\color{black}}{\thesubsubsection}{0.6em}{}
\titleformat{\paragraph}[runin]
  {\normalsize\bfseries\color{black}}{}{0pt}{}[.]
\titlespacing*{\section}{0pt}{2.2ex plus 0.6ex minus 0.2ex}{1.0ex}
\titlespacing*{\subsection}{0pt}{1.8ex plus 0.4ex minus 0.2ex}{0.7ex}
\titlespacing*{\paragraph}{0pt}{1.0ex plus 0.3ex minus 0.2ex}{0.6em}

\definecolor{navyblue}{RGB}{20, 55, 90}
\definecolor{midblue}{RGB}{40, 90, 135}
\definecolor{softblue}{RGB}{236, 243, 250}
\definecolor{accentteal}{RGB}{30, 120, 130}
\definecolor{softteal}{RGB}{235, 246, 247}
\definecolor{accentorange}{RGB}{190, 80, 25}
\definecolor{softorange}{RGB}{253, 244, 238}
\definecolor{darkgray}{RGB}{40, 45, 50}
\definecolor{midgray}{RGB}{100, 110, 120}
\definecolor{lightgray}{RGB}{246, 248, 250}
\definecolor{cardbg}{RGB}{255, 255, 255}
\definecolor{bordergray}{RGB}{205, 205, 205}

\fancypagestyle{paperstyle}{%
  \fancyhf{}
  \fancyhead[L]{\footnotesize\color{papergray}\textsc{Reading the News Without Hindsight}}
  \fancyhead[R]{\footnotesize\color{papergray}\thepage}
  \renewcommand{\headrulewidth}{0.35pt}
  \renewcommand{\headrule}{\hbox to\headwidth{\color{bordergray}\leaders\hrule height \headrulewidth\hfill}}
  
}
\AtBeginEnvironment{thebibliography}{\small\setlength{\bibsep}{0.45em}}

\newcommand{\topA}[1]{\cellcolor{accentteal!42}\textbf{#1}}
\newcommand{\topB}[1]{\cellcolor{accentteal!24}#1}
\newcommand{\topC}[1]{\cellcolor{accentteal!10}#1}
\newcommand{\shadekey}{\setlength{\fboxsep}{1.2pt}%
  Shading marks the best (\colorbox{accentteal!42}{\strut\textbf{darkest}}),
  second (\colorbox{accentteal!24}{\strut second}) and third
  (\colorbox{accentteal!10}{\strut third}) value of each column}
\newcolumntype{L}[1]{>{\raggedright\arraybackslash}p{#1}}
\newcommand{\tablenotes}[1]{%
  \par\vspace{0.45em}
  \begin{minipage}{\linewidth}\footnotesize\textit{Notes:} #1\end{minipage}}



\newcommand{\maintablehead}{%
\toprule
 & & \multicolumn{2}{c}{Forecast} & \multicolumn{4}{c}{Net Sharpe ratio} & \multicolumn{5}{c}{Buffered long-short, net} \\
\cmidrule(lr){3-4}\cmidrule(lr){5-8}\cmidrule(l){9-13}
Family & Feature set $\cdot$ \textit{model} & IC & $t$ & Dec. & Buf. & \makecell[c]{Dec.\\cw} & \makecell[c]{Buf.\\cw} & \makecell[c]{Ret.\\(\%)} & \makecell[c]{Vol.\\(\%)} & \makecell[c]{Max.\\DD} & \makecell[c]{Turn.\\(\%)} & \makecell[c]{$\alpha$\\$t$} \\
\midrule}
\newcommand{\maintablenotes}{%
\tablenotes{\shadekey. Upper block: the experiment with the best validation IC within each family; a row lists every family that picks it (Overall: all base-model experiments of the main design; Tone/emb.: the tone and embedding sets; 63\,d shocks: the sets with the 63-day window; Every set: all feature sets; Ranked targets: all models including the ranked-target fits; reader names: that reader's sets). Lower block: equal-weight combinations (Section~\ref{sec:methods:ensembles}). Dec., Buf.: decile and buffered long-short; cw: legs weighted by market value. Return, volatility, maximum drawdown (\%), turnover per period and the $t$ statistic of the three-factor alpha refer to the buffered equal-weighted portfolio.}}

\newcommand{\todo}[1]{\textcolor{gray}{\textit{[#1]}}}

\newcommand{\orcidicon}{%
  \href{https://orcid.org/0009-0007-4120-0921}{%
    \raisebox{-0.8pt}{\includegraphics[height=9pt]{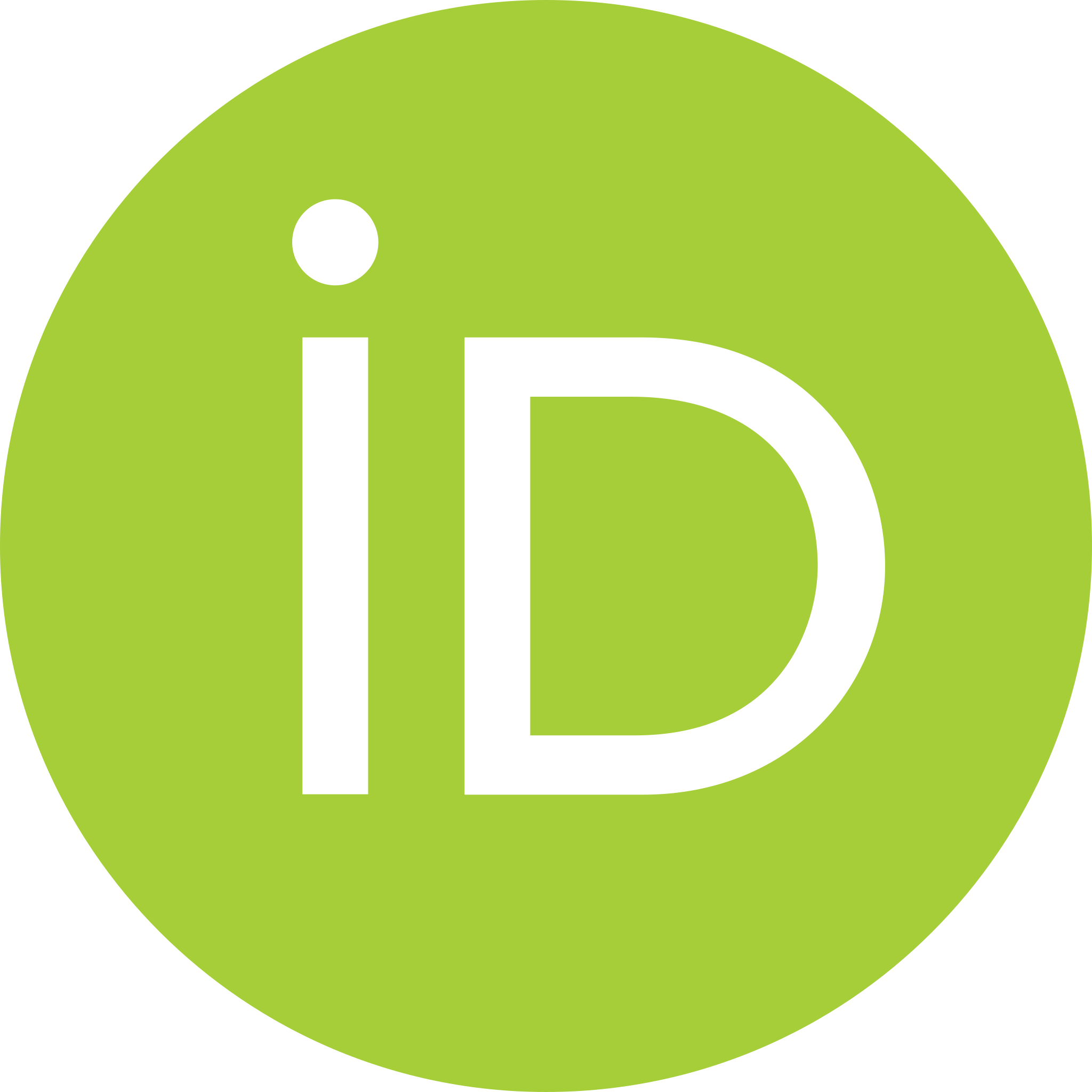}}}}

\renewcommand{\maketitle}{%
  \thispagestyle{empty}
  \vspace*{0.42in}
  \begin{center}
    {\fontsize{24}{29}\selectfont\bfseries
      Reading the News Without Hindsight:\\[0.12em]
      NoLBERT Topics and the\\[0.12em]
      Cross-Section of Stock Returns\par}
    \vspace{1.35em}
    {\large
      Kevin Foley\textsuperscript{*}\quad
      Jonathan Hartadi\textsuperscript{*}\\[0.35em]
      Shivesh Prakash\textsuperscript{*,\(\dagger\)}\,\orcidicon\quad
      Charishma Takkallapalli\textsuperscript{*}\quad
      Swapnil Vatsal\textsuperscript{*}\par}
    \vspace{0.75em}
    {\normalsize University of California, Berkeley\par}
    {\small Berkeley, California, USA\par}
    \vspace{0.85em}
    {\small
      \href{mailto:kevinfoley3@berkeley.edu}{kevinfoley3@berkeley.edu}
      \enspace\textperiodcentered\enspace
      \href{mailto:johartadi23@berkeley.edu}{johartadi23@berkeley.edu}\\[0.25em]
      \href{mailto:shivesh_prakash@berkeley.edu}{shivesh\_prakash@berkeley.edu}
      \enspace\textperiodcentered\enspace
      \href{mailto:charishma_t@berkeley.edu}{charishma\_t@berkeley.edu}
      \enspace\textperiodcentered\enspace
      \href{mailto:swapnil_vatsal@berkeley.edu}{swapnil\_vatsal@berkeley.edu}\par}
  \end{center}
  \vspace{1.45em}
  \begingroup
    \renewcommand{\thefootnote}{\fnsymbol{footnote}}%
    \footnotetext[1]{Equal contribution; authors listed alphabetically by last name.}%
    \footnotetext[2]{Corresponding author.}%
  \endgroup
}

\begin{document}
\maketitle

\begin{abstract}
\noindent
Language models that read financial news are usually trained on text from the same years used to test them, so a backtest can reward hindsight. We build news topics from NoLBERT, an encoder pretrained only on text published between 1976 and 1995, adapted into a sentence encoder by contrastive fine-tuning on passages of the same article, with no labels or returns. We compare it with LDA, MiniLM and FinBERT readers on 1.26 million articles from 2012 to 2024, convert each reader's topics into attention shocks and stock exposures, and forecast the returns of the 1,000 largest US stocks with eight models at three rebalancing frequencies. Models are fitted on 2013--2019, choices are made on 2020--2021, and 2022--2024 is held out. The experiment with the best validation score loses money at every frequency. Measuring attention shocks against the past 63 trading days instead of 5 raises the test rank IC of every reader. NoLBERT gives the most stable topics and forecasts about as well as LDA. An equal-weight combination of all readers and models earns a monthly net Sharpe ratio of 1.12, significant against zero before adjusting for the search but not against a price-only benchmark.
\end{abstract}

\vspace{0.8em}
\begin{center}
\small
\textbf{Keywords:}
narrative asset pricing; NoLBERT; look-ahead bias; contrastive fine-tuning;
financial news; return prediction
\end{center}
\vfill
\clearpage
\pagestyle{paperstyle}

\section{Introduction}
\label{sec:intro}

Financial news carries information about the economy, about firms and about the events that move prices, and large news archives \citep{dong2024fnspid} now make it possible to test whether that information forecasts returns. Transformer encoders can read such archives with some sense of context, placing two articles about the same event close together even when they share few words \citep{reimers2019sentencebert}. They also bring a problem into historical research. An encoder released in 2021 has read text written during most of any sample that ends before then, so the vector it gives a 2018 article already reflects what happened after 2018 \citep{kakhbod2025nolbert}. Holding out a test period for the return model does not remove this, because the hindsight sits in the features.

How much this matters depends on the model and the task. Language models can recall economic data from their training period \citep{lopezlira2025memorization}, and their forecasts of events inside that period can be biased upwards \citep{sarkar2024lookahead}. For news-based return prediction, \citet{he2025chronologically} find the bias to be modest. The same question applies to topic models and vocabularies estimated on the full sample.

Our earlier study \citep{prakash2026ssrn,foley2026wordcountscontexttopic} compared LDA with a frozen sentence transformer in the narrative asset-pricing framework of \citet{bybee2023narrative} and found the transformer ahead on both coherence and Sharpe ratio, though the difference in performance was not statistically significant. That design had weaknesses that made the result hard to carry out of sample: topic models and vocabularies were fitted on the full sample, the stock universe used information not yet available at the start of each year, some choices were made on the evaluation months, and portfolios were evaluated before trading costs. This paper repeats the comparison under a stricter protocol. Related theoretical work highlights how heterogeneous trading decisions and market structure affect trading activity and welfare \citet{back2023segmented}.

The central addition is a point-in-time encoder. NoLBERT \citep{kakhbod2025nolbert} was pretrained only on text published between 1976 and 1995, long before our news begins, but its raw output is not suited to comparing whole articles. We adapt it by contrastive fine-tuning on pairs of passages from the same article \citep[as in][]{gao2021simcse}, without returns or labels, cluster the adapted embeddings into topics, and compare these topics with LDA, MiniLM and a FinBERT topic classifier. All six readers label the same 1.26 million articles from 2012 to 2024 and are fitted on text up to 2019. Following \citet{bybee2023narrative}, we turn each reader's topics into daily attention, attention shocks and stock exposures to those shocks, add firm-level news and price characteristics, and forecast the relative returns of the 1,000 largest US stocks with eight models, from least squares and IPCA to tree ensembles and the explainable boosting machine (EBM) of \citet{bell2026glassbox}. The forecasts are traded under nine portfolio rules at monthly, biweekly and weekly frequencies, net of costs. Models are fitted on 2013--2019, every choice is made on 2020--2021, and 2022--2024 is used only for reporting.

The results come in four parts. First, the adapted NoLBERT gives the most stable and most even topics of the six readers and forecasts about as well as LDA, the strongest of the readers trained on recent text. We find no evidence that the point-in-time restriction costs forecasting power, although the test cannot rule out moderate differences between readers. Second, how attention is measured matters more than which reader measures it: comparing attention with its mean over the past 63 trading days instead of the past week raises the test rank IC of every reader at every frequency. Third, choosing one experiment on the validation period fails. The experiment with the best validation IC loses money at every frequency, and opening the EBM shows why. An equal-weight combination of all readers and models with the 63-day window earns net Sharpe ratios of 1.12 (decile) and 1.19 (buffered) monthly. These are significant against zero before any adjustment for the size of the search, but they do not beat a price-only ensemble, and no single experiment survives a false-discovery correction. Fourth, we document problems in the news and return data that can create spurious predictability, among them article bodies attached to the wrong headline, a news source collected after the fact, and missing returns that depend on the future.

Section~\ref{sec:related} reviews related work, Section~\ref{sec:data} describes the data, Section~\ref{sec:methods} the readers, the adaptation of NoLBERT and the forecasting design, and Section~\ref{sec:results} the results. Section~\ref{sec:discussion} discusses the limitations.

\section{Related Work}
\label{sec:related}

\paragraph{Text and asset prices} Research that turns documents into numbers and relates them to economic outcomes is surveyed by \citet{gentzkow2019text}. In finance, early work measured the tone of a daily news column and found that pessimism predicts lower market returns, later reversed, and higher trading volume \citep{tetlock2007giving}. \citet{ke2019predicting} learn the words that predict returns directly, so the text representation is supervised by the outcome. We keep returns out of every text model, since our comparison is between representations. \citet{bybee2023narrative} fit LDA to the \textit{Wall Street Journal}, turn the attention each topic receives into shocks, and use stocks' covariances with those shocks as characteristics in a factor model. Our pipeline follows theirs up to the exposures and then forecasts returns with a set of models. \citet{chen2022expected} use representations from large language models to predict the cross-section of returns, and we include their embedding regression as one of our feature sets.

A broader literature uses textual information to measure
economic conditions and predict financial outcomes.
\citet{baker2016measuring} construct a news-based
measure of economic policy uncertainty, while
\citet{manela2017news} and \citet{calomiris2019news}
show how information extracted from news can predict
market risk and returns. Related work applies
computational linguistics to central bank communication
\citep{hansen2018transparency}.
Our study contributes to this literature by examining
how alternative text representations, particularly
a historically pretrained language model, affect
the measurement of narrative attention and
out-of-sample return prediction.

Text-based measurement has spread across finance. \citet{kakhbod2026fedmind} extract latent dimensions from Federal Reserve records, \citet{kakhbod2025creative} and \citet{hochberg2026patents} build measures from patent text, \citet{kakhbod2025finfluencers} study financial content on social media, and \citet{chen2023process} study how process intangibles shape agency conflicts. These corpora are written by different authors and on different schedules from daily newswire copy, and none of them produces the daily attention series that the narrative framework needs.

\paragraph{Topic models and sentence encoders} LDA represents a document as a mixture of topics estimated from word counts, so word order carries no information \citep{blei2003lda}. Sentence encoders summarize a passage in one vector that can be compared with others by cosine similarity \citep{reimers2019sentencebert}. MiniLM, the encoder of our earlier study, is a small distilled model of this kind \citep{wang2020minilm}, and FinBERT adapts BERT to financial text \citep{huang2023finbert}. Our adaptation of NoLBERT uses the contrastive objective of \citet{gao2021simcse}, in which two views of the same text are pulled together and other texts in the batch are pushed apart. We judge topics partly by NPMI coherence \citep{roder2015coherence}, which depends on how each topic's top words are chosen, so we use one rule for every reader.

\paragraph{Look-ahead in language models} A model trained on text written during the sample can carry later information into the features of a backtest. \citet{lopezlira2025memorization} show that large language models recall economic and financial data from their training period, \citet{sarkar2024lookahead} document look-ahead bias in pretrained models used for forecasting, and \citet{he2025chronologically} train models on chronologically restricted text and find that the bias in news-based return prediction is modest. NoLBERT addresses the problem by pretraining only on text from 1976 to 1995 \citep{kakhbod2025nolbert}. A related concern is that generated text depends on the prompt and can carry systematic biases \citep{fedyk2026ai}, which is one reason we cluster fixed embeddings instead of asking a generative model to label topics.

\paragraph{Forecasting and evaluation} We use the machine-learning approach to return prediction of \citet{gu2020empirical} and include IPCA, in which characteristics drive time-varying factor loadings \citep{kelly2019characteristics}. Interpretable models have been used to study corporate bond returns \citep{bell2026glassbox}, portfolio choice \citep{bell2026alphaglass} and mortgage lending \citep{bell2025fairness}, and we use the EBM of \citet{bell2026glassbox} to see which features drive a forecast. Combining forecasts is often more reliable than selecting one \citep{rapach2010outofsample}. Because we run hundreds of experiments per frequency, we follow \citet{harvey2016cross} in treating single significant results with suspicion and report deflated Sharpe ratios \citep{bailey2014deflated} and tests of the best result in the search \citep{white2000reality,hansen2005test}.

\section{Data}
\label{sec:data}
\suppressfloats[t]

We combine financial news from 2012 to 2024 with the daily returns of the largest US stocks. The design follows our earlier study \citep{prakash2026ssrn,foley2026wordcountscontexttopic} with three changes: a longer news sample with repaired text, a complete return file, and a strict order in time. Topic models see text up to 2019, every model choice is made on 2020--2021, and results are reported for 2022--2024.

\subsection{News}
\label{sec:data:news}

The list of articles comes from FinMultiTime \citep{xu2025finmultitime}, a collection of Nasdaq.com articles filed under the ticker pages on which they appeared. The text comes from FNSPID \citep{dong2024fnspid}, which covers the same articles up to 2023. For 93 tickers FinMultiTime pairs the correct headline and date with the body of an unrelated article, often one written years later. This affects 7.3\% of the articles of 2012--2023, and left unrepaired it would leak later language into a vocabulary meant to end in 2019. FNSPID does not cover 2024, so for that year we use FinMultiTime's text where it is intact and the headline alone otherwise.

\begin{figure}[tbp]
\centering
\includegraphics[width=\linewidth]{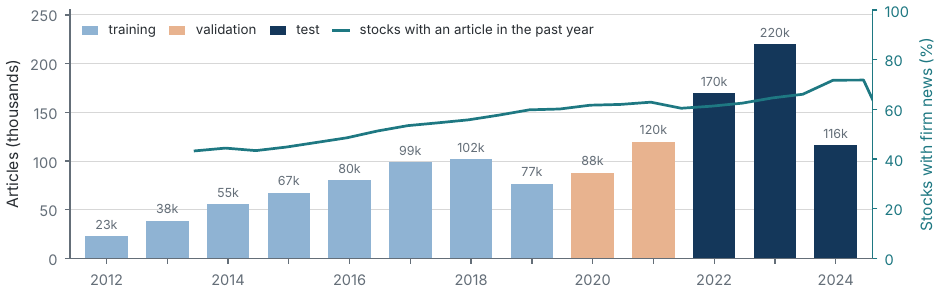}
\caption{Articles per year, coloured by the split of the return models, and the share of the stock universe with at least one article filed under its ticker in the previous 252 trading days (right axis).}
\label{fig:sample}
\end{figure}

We keep articles with a body of at least 400 characters, merge duplicates, and drop 16 December 2023, when the source holds a dump of tens of thousands of rows. This leaves 1{,}255{,}289 articles (Figure~\ref{fig:sample}). Every model reads the headline and the first 500 characters of the body. Article dates carry no time of day, so the news of day $d$ is first used at the close of day $d+1$. Coverage grows from a median of 83 articles per weekday in 2012 to 789 in 2023. To keep market-wide attention comparable across years, all topic models label the same random sample of at most 150 articles per day (529{,}657 articles) and are fitted on at most 60 per day dated up to 2019 (140{,}209 articles). The 2024 articles come from a second collection with fewer tickers and no headlines, so we also report 2022--2023 and 2024 separately.

Firm-level features use the articles filed under at most three tickers. The source was collected from ticker pages at the end of the sample and is not point-in-time. Companies acquired during the sample have no articles, and heavily covered firms keep only their most recent ones. We treat a firm with no article in the previous 126 trading days as not covered, which is the case for 50\%, 39\% and 40\% of stock-months in the training, validation and test periods.

\subsection{Returns}
\label{sec:data:returns}

Returns and market values come from the CRSP daily and monthly stock files \citep{crspstockdatabase}, and the Fama--French factors and risk-free rate from the Kenneth R.\ French Data Library \citep{frenchlibrary}. The universe consists of the 1{,}000 largest ordinary common shares (share codes 10 and 11 on NYSE, AMEX and Nasdaq) by market value at the end of each June from 2010 to 2024, and each list is used from that June until the next. We hold every daily return of the 2{,}062 stocks that enter a list, from 2011 to 2024. Portfolios are formed at the close of the last trading day of each month, fortnight or week. A stock enters a period if it is on the current list and has traded in the last five days, which gives about 984 stocks per period and panels of 141{,}761 stock-months, 306{,}746 stock-fortnights and 614{,}525 stock-weeks. A stock that stops trading during a holding period earns zero for the rest of it. The return file matters. From 2017 on, an earlier extract held daily returns only for stocks on the current calendar year's June list, so from January to June a stock about to leave the list had no recent returns, and such stocks went on to underperform by 3\% to 5\% a month. A tree model learned to find them, and the resulting weekly net Sharpe ratio of 1.56 on price features alone came from this gap. The complete file removes this, and no stock in our panels lacks a recent return.

\subsection{Sample Splits}
\label{sec:data:splits}

The training period runs from 2013 to 2019 (84 months, 182 fortnights, 365 weeks), validation from 2020 to 2021 (24, 52, 104) and the test period from 2022 to 2024 (36, 78, 156). A holding period belongs to the split in which it starts. Topic models, vocabularies and cluster centroids are fitted on text up to 2019 and then applied unchanged. Hyperparameters and the attention-shock window are chosen on the validation period only. Each forecaster is then refitted on 2013--2021 and its test forecasts come from that frozen model.

\section{Methods}
\label{sec:methods}

\subsection{Pipeline}
\label{sec:methods:overview}

\begin{figure}[tbp]
\centering
\resizebox{\linewidth}{!}{\input{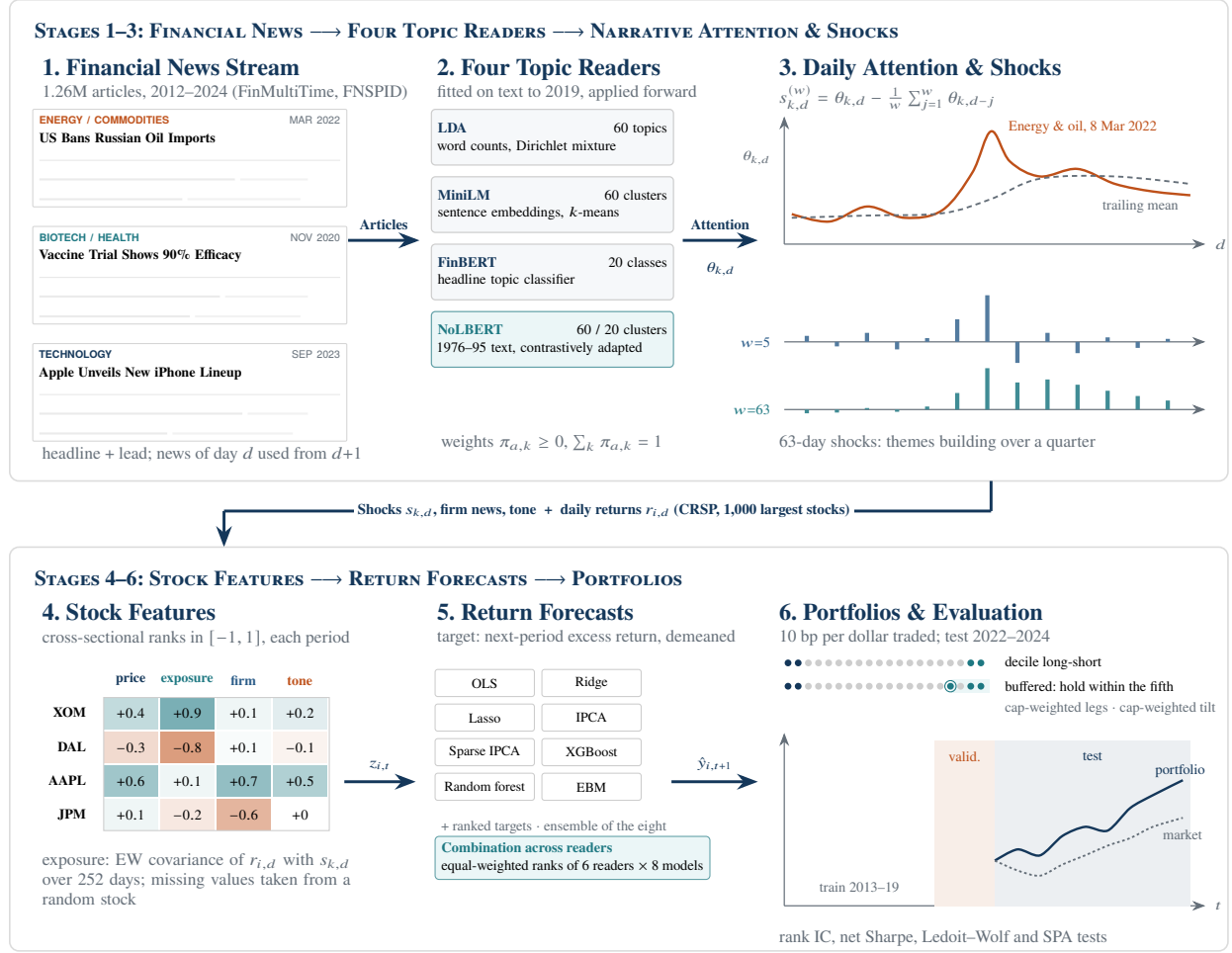}}
\caption{The pipeline. Each article is read by four families of topic readers. A day's article weights give each topic's attention, and its deviation from the mean of the previous $w$ trading days is the attention shock. Narrative exposures, firm news and price characteristics form the features; eight models and their combinations produce forecasts; nine rules turn them into portfolios.}
\label{fig:pipeline}
\end{figure}

Figure~\ref{fig:pipeline} shows the pipeline. A topic model, which we call a reader, gives every article a weight on each of its topics. Averaging the weights over a day's articles gives each topic's share of attention, and the gap between that share and its recent mean is the topic's attention shock. The covariance of a stock's returns with a topic's shocks over the past year is the stock's exposure to that narrative \citep{bybee2023narrative}. These exposures, the composition and tone of the firm's own news, and eight price characteristics are the features. Eight forecasting models turn them into forecasts of next period's relative return, and nine portfolio rules turn the forecasts into portfolios. Everything after the reader is identical across readers, so differences in the results come from the text representation.

Our earlier study compared LDA with a frozen sentence transformer inside a single Sparse IPCA pricing model \citep{prakash2026ssrn,foley2026wordcountscontexttopic}. Here we add a point-in-time language model adapted to financial news, use a set of forecasters in place of the single pricing model, and trade the forecasts at three rebalancing frequencies.

\subsection{Topic Readers}
\label{sec:methods:readers}

A reader maps article $a$ to weights $\pi_{a,k}\ge 0$ with $\sum_k \pi_{a,k}=1$ over its $K$ topics. All readers see the same text and label the same daily sample, and all topics are fitted on text up to 2019. We use six.

LDA \citep{blei2003lda} has $K=60$ topics, estimated by collapsed Gibbs sampling on a vocabulary of 10{,}000 unigrams and bigrams; topic proportions of later articles are inferred with the topic--word matrix held fixed. A second version, LDA (all articles), applies the same topics to every article instead of the daily sample. MiniLM is the reader of our earlier study \citep{prakash2026ssrn}: the \texttt{paraphrase-\allowbreak MiniLM-\allowbreak L3-\allowbreak v2} encoder \citep{reimers2019sentencebert,wang2020minilm} embeds each article, spherical $k$-means finds 60 clusters among the training embeddings, and an article's weights are a softmax of its cosine similarity to the centroids at temperature 0.07. FinBERT gives the probabilities of 20 fixed subjects from a FinBERT classifier applied to headlines \citep{huang2023finbert}. NoLBERT and NoLBERT-20 cluster the embeddings of our adapted NoLBERT encoder (Sections~\ref{sec:methods:nolbert} and~\ref{sec:methods:finetune}) into 60 and 20 topics in the same way as MiniLM.

Daily attention to topic $k$ is the mean weight over the sampled articles $\mathcal{A}_d$ of day $d$,
\begin{equation}
\theta_{k,d} \;=\; \frac{1}{|\mathcal{A}_d|}\sum_{a\in\mathcal{A}_d}\pi_{a,k}.
\label{eq:attention}
\end{equation}
Articles from non-trading days count towards the next trading day, and days with fewer than 20 articles have no reading.

\subsection{A Point-in-Time Encoder: NoLBERT}
\label{sec:methods:nolbert}

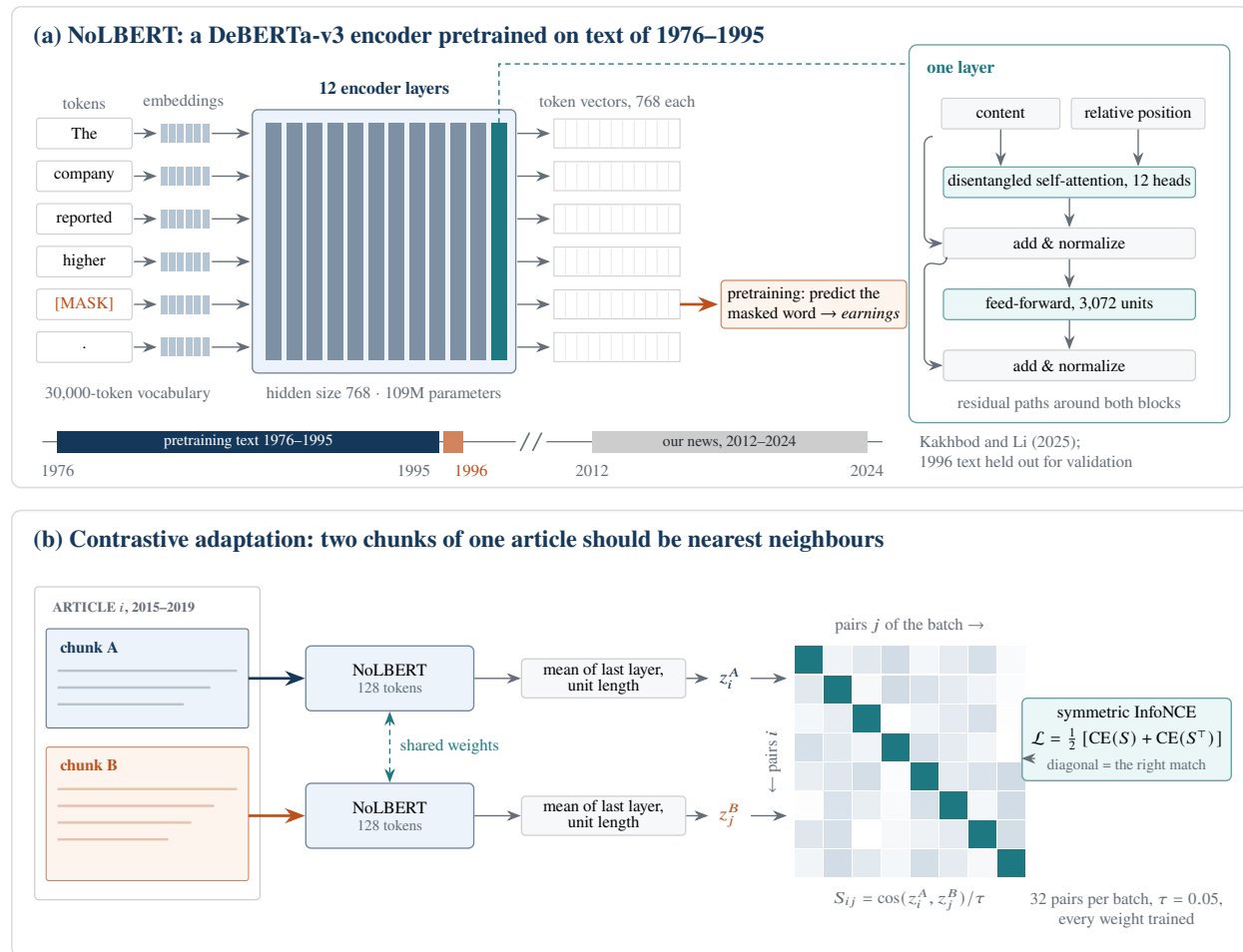
\begin{figure}[htbp]
\centering
\resizebox{\linewidth}{!}{
\begin{tikzpicture}[
    >=Stealth,
    font=\sffamily,
    panel/.style={draw=bordergray!70, fill=white, rounded corners=3.5pt, line width=0.6pt},
    header/.style={font=\bfseries\footnotesize\color{navyblue}},
    annot/.style={font=\fontsize{6.2}{7.4}\selectfont\color{midgray}},
    small/.style={font=\fontsize{6.4}{7.6}\selectfont},
    tok/.style={draw=bordergray, fill=white, rounded corners=1pt, line width=0.4pt, minimum width=1.25cm,
                minimum height=0.40cm, inner sep=1pt, font=\fontsize{6.2}{7}\selectfont},
    op/.style={draw=bordergray, fill=lightgray, rounded corners=1.5pt, line width=0.45pt, minimum width=3.3cm,
               minimum height=0.40cm, inner sep=1.5pt, font=\fontsize{6.2}{7}\selectfont},
    opk/.style={op, fill=softteal, draw=accentteal!70},
    flow/.style={->, line width=0.5pt, color=midgray},
    dataflow/.style={->, line width=1.0pt, color=navyblue}
]

\draw[panel] (0, 6.15) rectangle (16.2, 12.45);
\node[header, anchor=north west] at (0.15, 12.33) {(a) NoLBERT: a DeBERTa-v3 encoder pretrained on text of 1976--1995};

\foreach \w [count=\i] in {The, company, reported, higher, {\color{accentorange}[MASK]}, .} {
  \pgfmathsetmacro{\y}{11.35 - 0.56*\i}
  \node[tok] (t\i) at (0.95, \y) {\w};
  \foreach \c in {0,...,5} {
     \pgfmathsetmacro{\x}{1.95 + 0.11*\c}
     \fill[midblue!35] (\x, \y-0.12) rectangle ++(0.08, 0.24);
  }
  \draw[flow] (1.6, \y) -- (1.9, \y);
  \draw[flow] (2.62, \y) -- (3.12, \y);
  \draw[flow] (6.62, \y) -- (7.05, \y);
  \draw[draw=bordergray, fill=white, line width=0.35pt] (7.1, \y-0.19) rectangle (8.75, \y+0.19);
  \foreach \c in {1,...,10} { \draw[bordergray!45, line width=0.25pt] (7.1+0.15*\c, \y-0.17) -- (7.1+0.15*\c, \y+0.17); }
}
\node[annot, anchor=south] at (0.95, 10.97) {tokens};
\node[annot, anchor=south] at (2.25, 10.97) {embeddings};
\node[annot, anchor=south] at (7.93, 10.97) {token vectors, 768 each};
\node[annot, anchor=north west] at (0.3, 7.6) {30{,}000-token vocabulary};

\draw[draw=navyblue!60, fill=softblue, rounded corners=2pt, line width=0.5pt] (3.15, 7.65) rectangle (6.6, 11.1);
\foreach \j in {0,...,11} {
  \pgfmathsetmacro{\x}{3.33 + 0.268*\j}
  \ifnum\j=11
    \fill[accentteal] (\x, 7.82) rectangle ++(0.2, 3.1);
  \else
    \fill[navyblue!55] (\x, 7.82) rectangle ++(0.2, 3.1);
  \fi
}
\node[font=\bfseries\fontsize{6.6}{7.6}\selectfont, color=navyblue, anchor=south] at (4.875, 11.13) {12 encoder layers};
\node[annot, anchor=north] at (4.875, 7.6) {hidden size 768 $\cdot$ 109M parameters};

\draw[dataflow, color=accentorange] (8.75, 8.55) -- (9.25, 8.55);
\node[draw=accentorange!70, fill=softorange, rounded corners=1.5pt, line width=0.45pt, align=left,
      font=\fontsize{6.2}{7.4}\selectfont, anchor=west, inner sep=2.5pt] at (9.28, 8.55)
      {pretraining: predict the\\masked word $\rightarrow$ \textit{earnings}};

\draw[accentteal, dash pattern=on 2pt off 1.5pt, line width=0.5pt] (6.38, 10.92) -- (6.38, 11.7) -- (11.75, 11.7);
\draw[draw=accentteal!70, fill=white, rounded corners=3pt, line width=0.5pt] (11.75, 7.0) rectangle (15.95, 11.95);
\node[font=\bfseries\fontsize{6.6}{7.6}\selectfont, color=accentteal, anchor=north west] at (11.85, 11.88) {one layer};
\node[tok, minimum width=1.55cm, fill=lightgray] (cnt) at (12.95, 11.05) {content};
\node[tok, minimum width=1.75cm, fill=lightgray] (pos) at (14.75, 11.05) {relative position};
\node[opk] (att) at (13.85, 10.15) {disentangled self-attention, 12 heads};
\node[op]  (n1)  at (13.85, 9.35) {add \& normalize};
\node[opk] (ff)  at (13.85, 8.55) {feed-forward, 3{,}072 units};
\node[op]  (n2)  at (13.85, 7.75) {add \& normalize};
\draw[flow] (cnt.south) -- (att.north -| cnt.south);
\draw[flow] (pos.south) -- (att.north -| pos.south);
\draw[flow] (att) -- (n1); \draw[flow] (n1) -- (ff); \draw[flow] (ff) -- (n2);
\draw[flow, rounded corners=3pt] (12.05, 10.75) -- (11.95, 10.75) -- (11.95, 9.35) -- (n1.west);
\draw[flow, rounded corners=3pt] (n1.south west) ++(0.05,0) -- ++(0,-0.08) -- (11.95, 9.1) -- (11.95, 7.75) -- (n2.west);
\node[annot, anchor=north] at (13.85, 7.47) {residual paths around both blocks};

\begin{scope}[shift={(0, 6.75)}]
  \draw[midgray, line width=0.45pt] (0.4, 0) -- (6.6, 0);
  \draw[midgray, line width=0.45pt] (7.0, 0) -- (11.4, 0);
  \draw[midgray, line width=0.6pt] (6.66, -0.12) -- (6.78, 0.12);
  \draw[midgray, line width=0.6pt] (6.82, -0.12) -- (6.94, 0.12);
  \fill[navyblue] (0.6, -0.14) rectangle (5.6, 0.14);
  \node[font=\fontsize{5.8}{6.6}\selectfont, color=white] at (3.1, 0) {pretraining text 1976--1995};
  \fill[accentorange!70] (5.65, -0.14) rectangle (5.9, 0.14);

  \fill[bordergray] (7.6, -0.14) rectangle (11.2, 0.14);
  \node[font=\fontsize{5.8}{6.6}\selectfont, color=darkgray] at (9.4, 0) {our news, 2012--2024};
  \node[annot, anchor=north] at (0.6, -0.16) {1976};
  \node[annot, anchor=north east] at (5.62, -0.16) {1995};
  \node[annot, anchor=north west, color=accentorange] at (5.66, -0.16) {1996};
  \node[annot, anchor=north] at (7.6, -0.16) {2012};
  \node[annot, anchor=north] at (11.2, -0.16) {2024};
\end{scope}
\node[annot, anchor=west, align=left] at (11.75, 6.62) {\citet{kakhbod2025nolbert};\\1996 text held out for validation};

\draw[panel] (0, 0) rectangle (16.2, 5.85);
\node[header, anchor=north west] at (0.15, 5.73) {(b) Contrastive adaptation: two chunks of one article should be nearest neighbours};

\draw[draw=bordergray, fill=white, rounded corners=1.5pt, line width=0.45pt] (0.3, 0.75) rectangle (3.25, 4.85);
\node[font=\fontsize{5.2}{6}\selectfont\bfseries, color=midgray, anchor=north west] at (0.4, 4.78) {ARTICLE $i$, 2015--2019};
\draw[draw=navyblue!70, fill=softblue, rounded corners=1pt, line width=0.4pt] (0.45, 3.0) rectangle (3.1, 4.3);
\node[font=\fontsize{6}{7}\selectfont\bfseries, color=navyblue, anchor=north west] at (0.5, 4.27) {chunk A};
\foreach \k in {0,1,2} { \draw[navyblue!30, line width=0.9pt] (0.6, 3.75-0.22*\k) -- (2.95-0.35*\k, 3.75-0.22*\k); }
\draw[draw=accentorange!70, fill=softorange, rounded corners=1pt, line width=0.4pt] (0.45, 1.0) rectangle (3.1, 2.75);
\node[font=\fontsize{6}{7}\selectfont\bfseries, color=accentorange, anchor=north west] at (0.5, 2.72) {chunk B};
\foreach \k in {0,1,2,3} { \draw[accentorange!35, line width=0.9pt] (0.6, 2.2-0.22*\k) -- (2.95-0.3*\k, 2.2-0.22*\k); }

\node[draw=navyblue!60, fill=softblue, rounded corners=2pt, line width=0.5pt, minimum width=2.2cm, minimum height=0.85cm,
      font=\fontsize{6.4}{7.4}\selectfont, align=center] (ea) at (4.95, 3.65) {NoLBERT\\[-1pt]{\color{midgray}\fontsize{5.6}{6.4}\selectfont 128 tokens}};
\node[draw=navyblue!60, fill=softblue, rounded corners=2pt, line width=0.5pt, minimum width=2.2cm, minimum height=0.85cm,
      font=\fontsize{6.4}{7.4}\selectfont, align=center] (eb) at (4.95, 1.85) {NoLBERT\\[-1pt]{\color{midgray}\fontsize{5.6}{6.4}\selectfont 128 tokens}};
\draw[<->, accentteal, line width=0.5pt, dash pattern=on 2pt off 1.5pt] (ea.south) -- node[right, annot, color=accentteal] {shared weights} (eb.north);
\draw[dataflow] (3.1, 3.65) -- (ea.west);
\draw[dataflow, color=accentorange] (3.1, 1.85) -- (eb.west);

\node[op, minimum width=2.15cm, align=center] (pa) at (7.75, 3.65) {mean of last layer,\\[-1pt] unit length};
\node[op, minimum width=2.15cm, align=center] (pb) at (7.75, 1.85) {mean of last layer,\\[-1pt] unit length};
\draw[flow] (ea) -- (pa); \draw[flow] (eb) -- (pb);
\node[font=\fontsize{6.6}{7.6}\selectfont, color=navyblue] (za) at (9.4, 3.65) {$z^A_i$};
\node[font=\fontsize{6.6}{7.6}\selectfont, color=accentorange] (zb) at (9.4, 1.85) {$z^B_j$};
\draw[flow] (pa) -- (za); \draw[flow] (pb) -- (zb);

\begin{scope}[shift={(10.25, 1.05)}]
  \foreach \r in {0,...,7} {
    \foreach \c in {0,...,7} {
      \pgfmathsetmacro{\v}{int(mod(37*\r + 53*\c + 11*\r*\c, 23))}
      \ifnum\r=\c
        \fill[accentteal] (0.38*\c, 2.66-0.38*\r) rectangle ++(0.36, 0.36);
      \else
        \fill[midblue!\v] (0.38*\c, 2.66-0.38*\r) rectangle ++(0.36, 0.36);
      \fi
    }
  }
  \node[annot, anchor=south] at (1.51, 3.05) {pairs $j$ of the batch $\rightarrow$};
  \node[annot, anchor=south, rotate=90] at (-0.05, 1.51) {$\leftarrow$ pairs $i$};
  \node[annot, anchor=north] at (1.51, -0.02) {$S_{ij}=\cos(z^A_i, z^B_j)/\tau$};
\end{scope}
\draw[flow] (za.east) -- (10.15, 3.65);
\draw[flow] (zb.east) -- (10.15, 1.85);

\node[draw=accentteal!70, fill=softteal, rounded corners=2pt, line width=0.5pt, align=center,
      font=\fontsize{6.4}{7.6}\selectfont, minimum width=2.55cm, inner sep=3pt] (loss) at (14.6, 2.85)
      {symmetric InfoNCE\\[2pt] $\mathcal{L}=\tfrac12\left[\mathrm{CE}(S)+\mathrm{CE}(S^{\top})\right]$\\[2pt]
       {\color{midgray}\fontsize{5.6}{6.6}\selectfont diagonal = the right match}};
\draw[flow] (13.4, 2.6) -- (loss.west |- 13.4, 2.6);
\node[annot, anchor=south, align=center] at (14.6, 0.25) {32 pairs per batch, $\tau=0.05$,\\every weight trained};
\end{tikzpicture}}
\caption{NoLBERT and our adaptation. (a) The pretrained encoder of \citet{kakhbod2025nolbert}: a DeBERTa-v3 base model whose twelve layers combine content and relative-position attention, trained to predict masked words in English text of 1976--1995. (b) Contrastive adaptation: two passages of the same article pass through the shared encoder, are averaged over tokens and normalized, and the symmetric InfoNCE loss rewards placing each passage closest to its partner among the 32 pairs of a batch.}
\label{fig:nolbert}
\end{figure}

Two of our readers are built on encoders trained on text from inside our sample. MiniLM was released in 2021, and the FinBERT topic classifier was fine-tuned on financial text released in 2022, which falls in our test period. LDA is fitted on our own articles up to 2019 and has not read later text. NoLBERT \citep{kakhbod2025nolbert} offers a pretrained encoder without this problem (Figure~\ref{fig:nolbert}a). It is a DeBERTa-v3 base model with twelve layers, hidden size 768, twelve attention heads and about 109 million parameters, trained to predict masked words in text published between 1976 and 1995, with 1996 held out for validation. Its tokenizer, with 30,000 word pieces, was also trained on that corpus, so no later text enters through the vocabulary, and position enters only through relative attention. \citet{kakhbod2025nolbert} report that the model scores at least as well as FinBERT on a subset of the GLUE benchmark and that its masked-word probabilities favour 1976--1995 phrasing over both later and earlier phrasing.

The restriction has a cost. A vocabulary built on older text splits current financial language into more pieces, 1.49 tokens per word on our articles against 1.43 for MiniLM, so the same 64-token window covers less of each article. A larger problem is that the pretrained model has no sentence-level objective. When we average its token vectors with padding masked, two unrelated articles from our test period have a mean cosine similarity of 0.53, against 0.59 for a headline and its own lead. Vectors this close together are hard to cluster into 60 topics, which is why we adapt the encoder before using it.

\subsection{Adapting NoLBERT by Contrastive Fine-Tuning}
\label{sec:methods:finetune}

We fine-tune NoLBERT so that passages from the same article end up close together, following the unsupervised objective of \citet{gao2021simcse}. No head is added and no labels are used; the article boundary is the only supervision. There is no labelled topic assignment for 1.26 million articles, and a label set chosen by us would fix the topics in advance.

\paragraph{Pairs} We split each article published up to the end of 2019, the end of our training period, into passages, drop passages that occur verbatim in more than one article (mostly publisher boilerplate), and draw one pair of different passages per article, so that an article's partner passage never appears as a negative for another pair in the same batch. This gives 230,931 pairs, of which 2,048 are held out. Passages are cut at 128 tokens. No returns, prices or text from the validation and test periods enter at this stage.

\paragraph{Objective} A passage's vector $z$ is the mean of its last-layer token vectors, scaled to unit length. The first-position vector is not an option, since DeBERTa was never trained with a sentence-level task on that position. For a batch of $B=32$ pairs with vectors $a_i$ and $b_i$, the loss is the symmetric InfoNCE
\begin{equation}
\mathcal{L} = -\frac{1}{2B}\sum_{i=1}^{B}\left[\log\frac{e^{a_i^{\top}b_i/\tau}}{\sum_{j}e^{a_i^{\top}b_j/\tau}} + \log\frac{e^{a_i^{\top}b_i/\tau}}{\sum_{j}e^{a_j^{\top}b_i/\tau}}\right], \qquad \tau = 0.05,
\label{eq:infonce}
\end{equation}
which pulls the two passages of an article together and pushes the other 31 articles of the batch away. The second force is what spreads the vectors apart, so we apply no whitening or other correction afterwards.

\paragraph{Estimation} All weights are trained with AdamW (learning rate $2\times 10^{-5}$, weight decay 0.01, linear warm-up then cosine decay, gradient clipping at 1) in bfloat16, since DeBERTa's relative-position terms overflow in half precision. The run was limited to 150 minutes on one RTX 3060 GPU and completed 6,205 steps, or 198,560 pairs, about 0.87 of a pass over the training pairs. In this run the token average also included the padding positions; the embeddings used for the topics average over actual tokens only.

\begin{figure}[htbp]
\centering
\includegraphics[width=\linewidth]{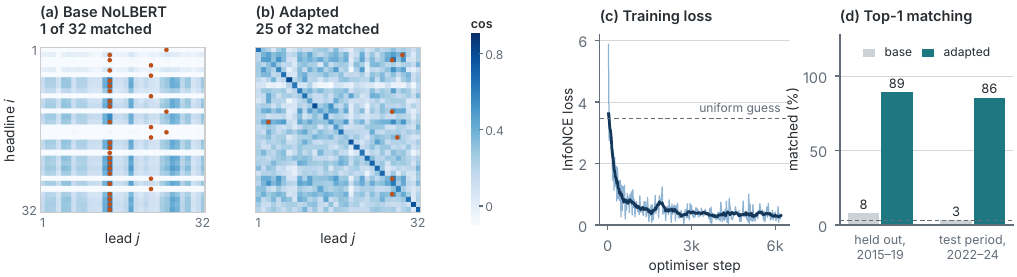}
\caption{What the adaptation changes. (a, b) Cosine similarity between the headline and the lead of 32 articles from 2022--2024, computed with the settings of the adaptation run; orange dots mark headlines whose most similar lead belongs to another article. Over 512 such pairs, matching pairs have a mean similarity of 0.05 under the base encoder and 0.56 after adaptation, against 0.04 and 0.14 for non-matching pairs. (c) Training loss. (d) Share of passages whose nearest neighbour among the 32 candidates of a batch is their partner, on the 2,048 held-out pairs of the adaptation run and on the 512 pairs from 2022--2024.}
\label{fig:finetune}
\end{figure}

\paragraph{Result} The loss falls from about 5.9 to about 0.3, well below $\log 32 = 3.47$, the loss of a model whose similarities carry no information (Figure~\ref{fig:finetune}c). On the 2,048 held-out pairs, the share of passages whose nearest neighbour among the 32 candidates of a batch is their partner rises from 0.08 to 0.89, against a chance level of 0.03. On 512 headline-and-lead pairs from 2022--2024, which the encoder never saw, it is 0.86 (Figure~\ref{fig:finetune}). To choose the pooling we clustered 11,988 articles into 60 topics with several candidate representations and scored the topics by NPMI. The mean of the last layer without whitening scored 0.343, against 0.268 to 0.282 for averages of the top four layers with or without whitening and 0.308 for MiniLM. The NoLBERT and MiniLM partitions of that sample agree only partly (adjusted mutual information 0.45).

\paragraph{From embeddings to topics} After adaptation the NoLBERT readers are built exactly like MiniLM (Section~\ref{sec:methods:readers}), with 60 centroids for NoLBERT and 20 for NoLBERT-20. One difference from MiniLM remains: NoLBERT is adapted to our corpus and MiniLM is used as published, so a comparison of the two mixes the effect of the pretraining period with that of the adaptation.

\subsection{Attention Shocks and Exposures}
\label{sec:methods:attention}

Following \citet{bybee2023narrative}, we work with surprises in attention. The shock to topic $k$ on day $d$, measured against the previous $w$ trading days, is
\begin{equation}
s^{(w)}_{k,d} \;=\; \theta_{k,d} \;-\; \frac{1}{w}\sum_{j=1}^{w}\theta_{k,d-j},
\label{eq:shock}
\end{equation}
made orthogonal to the market return within each estimation window. A stock's exposure to the narrative is the exponentially weighted covariance of its daily return with the shock over the 252 trading days before the formation date (half-life 60 days, at least 120 returns),
\begin{equation}
b^{(w)}_{i,k,t} \;=\; \widehat{\operatorname{Cov}}^{\,\mathrm{EW}}_{d< t}\!\left(r_{i,d},\, s^{(w)}_{k,d}\right).
\label{eq:exposure}
\end{equation}

The main design uses $w=5$, which measures the news of the past week. We also chose $w$ on the validation period with a rule written down beforehand. Exposures of every reader were rebuilt for $w\in\{1,5,21,63\}$; OLS, ridge and lasso were fitted on the training period for each reader and frequency; and the window with the highest average validation rank IC was adopted if it also beat $w=5$ for at least four of the six readers. The rule picked $w=63$ at all three frequencies. Feature sets built with it are marked ``63\,d''.

\subsection{Features}
\label{sec:methods:features}

Table~\ref{tab:blocks} lists the feature blocks. Every feature enters as its cross-sectional rank at the formation date, scaled to $[-1,1]$. A missing value can depend on the future in our data, since a firm without news is more often one that was later acquired. So a stock with a missing value in a block receives the full block of a randomly drawn stock of the same date before ranking, and keeps its true return. This applies to 0.3\% to 1.2\% of stock-periods for prices, under 1\% for narrative exposures, and 39\% to 50\% for firm news and tone.

\begin{table}[tbp]
\centering
\small
\caption{Feature blocks. $K$ is the number of topics of the reader.}
\label{tab:blocks}
\setlength{\tabcolsep}{5pt}
\renewcommand{\arraystretch}{1.1}
\begin{tabular}{@{}L{3.1cm}r L{10.4cm}@{}}
\toprule
Block & Number & Definition \\
\midrule
Price & 8 & Market beta and idiosyncratic volatility (252 days); 21-day volatility and maximum return; momentum ($t-251$ to $t-21$); 21-day and 5-day reversal; log market value \\
Narrative exposures & $K$ & Equation~\eqref{eq:exposure} for each topic, with $w=5$ or $w=63$ \\
Firm news & $1+K$ & The firm's own articles over 21 days: $\log(1+\text{count})$ and their share in each topic \\
Placebo exposures & $K$ & LDA exposures with the shocks moved back by 126 trading days \\
FinBERT tone & 8 & Headline tone \citep{araci2019finbert} over 1, 5, 21 and 252 days, its surprise, article counts and abnormal news volume \\
Dictionary tone & 5 & Loughran--McDonald tone \citep{loughran2011liability} over 1, 5, 21 and 252 days and its surprise \\
Headline embeddings & 64 & Principal components of the firm's mean FinBERT headline vector over the last holding period \citep{chen2022expected} \\
News $\times$ price & 6 & Recent returns of stocks with and without news, and returns times tone \\
\bottomrule
\end{tabular}
\end{table}

The main design has 26 feature sets per frequency: price alone; for each reader the narrative exposures, the firm news, both together (``text'') and both with price (``price + text''); and the placebo. A second group adds the narrative, text and price-and-text sets of every reader with $w=63$, and seven sets built on the tone and embeddings of the news, alone and with price.

\subsection{Forecasting Models}
\label{sec:methods:prediction}

The target is the excess return of stock $i$ over the holding period minus its cross-sectional mean,
\begin{equation}
y_{i,t+1} \;=\; \bigl(r_{i,t+1}-r^{f}_{t+1}\bigr) \;-\; \frac{1}{N_t}\sum_{j=1}^{N_t}\bigl(r_{j,t+1}-r^{f}_{t+1}\bigr),
\label{eq:target}
\end{equation}
so the models forecast which stocks will beat the others. Returns are not winsorized. Six of the models are also fitted on the cross-sectional rank of $y_{i,t+1}$ to limit the influence of extreme returns on a squared-error fit.

Table~\ref{tab:models} lists the eight base models, close to the set compared by \citet{gu2020empirical}: three linear regressions; IPCA, in which characteristics drive time-varying factor loadings \citep{kelly2019characteristics}, and the group-sparse IPCA of our earlier study; two tree ensembles, a random forest and XGBoost \citep{chen2016xgboost}; and the explainable boosting machine \citep[EBM;][]{lou2013accurate}, an additive model of single-feature and pairwise shape functions that \citet{bell2026glassbox} use as a glass-box model of asset returns.

\begin{table}[tbp]
\centering
\small
\caption{Forecasting models and the settings tried on the validation period.}
\label{tab:models}
\setlength{\tabcolsep}{5pt}
\renewcommand{\arraystretch}{1.1}
\begin{tabular}{@{}L{2.3cm}L{8.8cm}L{3.5cm}@{}}
\toprule
Model & Specification & Settings tried \\
\midrule
OLS & Least squares & --- \\
Ridge & $\ell_2$ penalty & 25 penalties \\
Lasso & $\ell_1$ penalty & 40 penalties \\
IPCA & $y_{i,t+1}=z_{i,t}'\Gamma f_{t+1}+\varepsilon$, forecast $z_{i,t}'\Gamma\bar f$ & 1, 2, 3 or 5 factors \\
Sparse IPCA & IPCA with a group penalty on the rows of $\Gamma$ & 1 or 3 factors, 3 penalties \\
XGBoost & Histogram trees, learning rate 0.05, early stopping on validation & Depth 1, 2, 3 or 5 \\
Random forest & 100 trees, at least 50 observations per leaf & Depth 3, 6 or 10 \\
EBM & Boosted one-feature step functions, then pairwise terms; early stopping on validation & 0 or 10 pairs \\
\bottomrule
\end{tabular}
\end{table}

Every setting of a model is fitted on the training period and scored by its mean squared error on the validation period. We keep the most regularized setting whose validation error is within $10^{-5}$ of the best, refit it on 2013--2021 and forecast the test period without further estimation. The EBM is the exception. Its main effects and pairs stop boosting at different rounds, so a refit with a fixed number of rounds would not reproduce the chosen model, and we keep the model fitted on the training period. This gives 564 experiments per frequency.

\subsection{Combinations}
\label{sec:methods:ensembles}

Forecasts are combined by averaging their within-period ranks with equal weights, which requires no estimation \citep{rapach2010outofsample}. The ensemble of a feature set averages its eight base models. The combination across readers averages the ensembles of the six readers for one kind of feature set (price + text, or text alone) and one shock window, so it pools 48 fitted models. The price ensemble, the ensemble on price alone, is the benchmark without text. We also average the price ensemble with the single tone signal that has the best validation IC, and the ten base-model experiments with the best validation IC.

For comparison we report, within each family of experiments (a kind of feature set, a reader, the 63-day sets, the ensembles and so on), the experiment with the highest validation IC. The ``overall pick'' is the best of all base-model experiments of the main design.

\subsection{Portfolios}
\label{sec:methods:portfolios}

Portfolios are formed at the close of each formation date and held to the next. The decile long-short buys the top tenth of forecasts and sells the bottom tenth with equal weights. The buffered long-short enters a stock in the top (bottom) tenth and holds it until it leaves the top (bottom) fifth, following the hold-out bands of \citet{novymarx2016taxonomy}. Both also appear with legs weighted by market value. The other rules are a rank-weighted portfolio of all stocks, dollar-neutral mean-variance weights under a five-factor covariance, the top decile alone with equal or market weights, and a market-weighted portfolio tilted towards the forecast; the last three are measured against the universe. Turnover is half the sum of absolute weight changes, and trading costs 10 basis points per dollar traded.

\subsection{Evaluation}
\label{sec:methods:evaluation}

The rank information coefficient (IC) of a period is the Spearman correlation across stocks between forecast and target. We report its mean, multiplied by 100, with a Newey--West $t$ statistic \citep{newey1987simple} judged against the fixed-$b$ critical values of \citet{kiefer2005new}, which are 2.29, 2.11 and 2.06 for the 36, 78 and 156 test periods. Sharpe ratios are annualized and net of costs. For portfolios with fixed forecasts we test the mean return against zero, report the probabilistic and deflated Sharpe ratios \citep{bailey2014deflated}, and compare two Sharpe ratios with the HAC test of \citet{ledoit2008robust} and a paired stationary bootstrap. Across all experiments of a frequency we run White's reality check \citep{white2000reality} and Hansen's test of superior predictive ability \citep{hansen2005test}. Paired comparisons match experiments that differ in one choice only and report the mean difference in IC and net return with Newey--West $t$ statistics.

\section{Results}
\label{sec:results}

All results are for the test period, 2022--2024, unless labelled otherwise. ICs are multiplied by 100 and Sharpe ratios are annualized and net of costs. In every table the three best values of a column are shaded. Figures and tables numbered A1, A2 and so on are in the appendix.

\subsection{Choosing One Experiment on Validation Fails}
\label{sec:results:validation}

\begin{figure}[tbp]
\centering
\includegraphics[width=\linewidth]{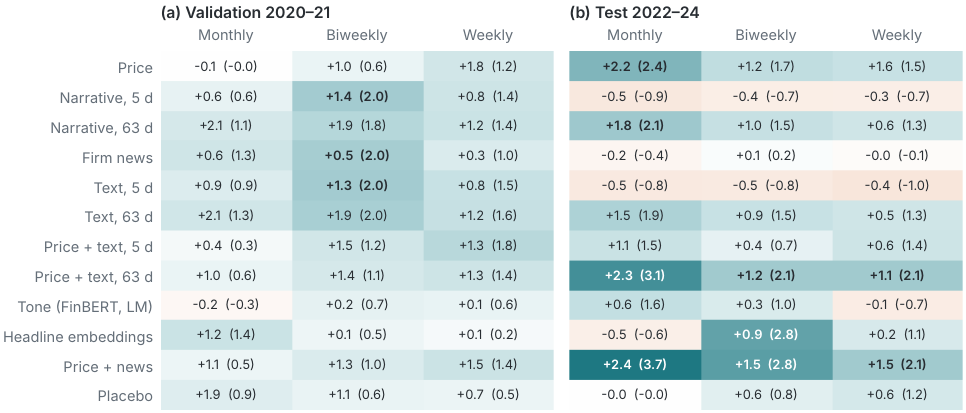}
\caption{Average IC and its $t$ statistic of the base models by kind of feature set, (a) in the validation period and (b) in the test period. Teal cells are positive, orange negative.}
\label{fig:landscape}
\end{figure}

Figure~\ref{fig:landscape} averages the IC of the eight base models over each kind of feature set, in the validation and the test period. The two periods disagree. In 2020--2021 the 5-day narrative and text sets and the firm's own news reach $t$ statistics of 2.0 at the biweekly frequency, and price alone looks weak. In 2022--2024 only the sets that contain price characteristics (2.2 monthly for price alone, $t=2.4$) or 63-day shocks have a positive average IC at every frequency. The 5-day narrative exposures are negative at every frequency, and firm news and tone are close to zero. The full numbers are in Table~\ref{tab:kinds}.

A rule that picks one experiment out of hundreds suffers most from this. The experiment with the best validation IC, the overall pick, is an EBM at every frequency, on price and LDA text monthly and weekly and on FinBERT text biweekly. Its test Sharpe ratios are $-0.23$, $-0.27$ and $-1.08$. Figure~\ref{fig:models} shows that the EBM has the highest validation IC of the base models at every frequency (2.4, 2.0 and 1.7) and a test IC close to zero. It stops boosting where the validation error is smallest, so validation flatters it more than any other model. The base models are within $-0.9$ and $0.6$ of each other in test IC. Models fitted on ranked returns do better, and XGBoost on ranks reaches 1.6, 1.0 and 1.2 ($t=1.8$, 1.5 and 2.6; Table~\ref{tab:models-results}).

\begin{figure}[tbp]
\centering
\includegraphics[width=\linewidth]{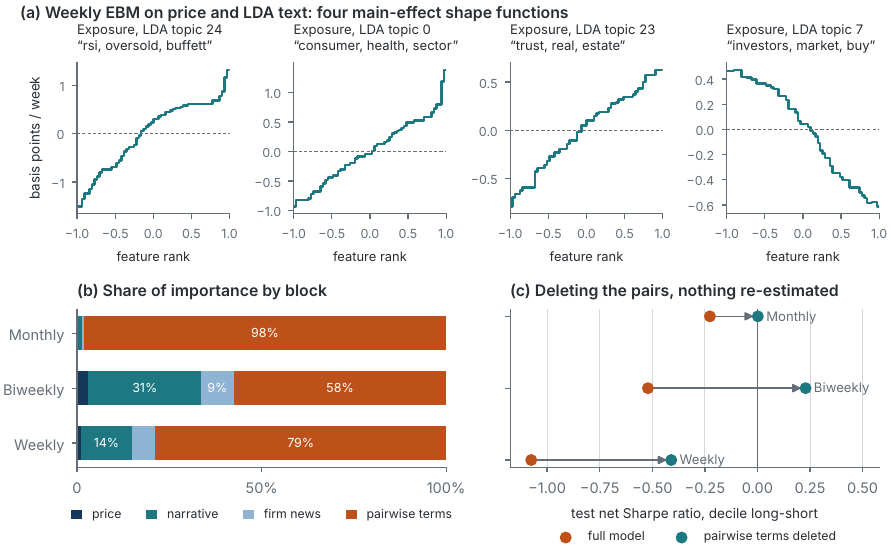}
\caption{The EBM on price and LDA text. (a) Four single-feature shape functions of the weekly model, for exposures to four LDA topics labelled by their top words. (b) Share of the mean absolute contribution by block. (c) Test Sharpe ratio of the decile long-short portfolio with and without the pairwise terms.}
\label{fig:ebm}
\end{figure}

\paragraph{Why the EBM fails} Because the EBM is additive, its forecasts split exactly into contributions of single features and of feature pairs \citep{bell2026glassbox}. We open the price-and-LDA-text EBM with the best validation IC at each frequency, which is also the overall pick monthly and weekly. Its single-feature effects look sensible: the forecast rises with the exposure to some narratives and falls with others, by one to three basis points a week across the cross-section (Figure~\ref{fig:ebm}a). But validation chose ten pairwise terms at every frequency, and they carry 98\%, 58\% and 79\% of the model (Figure~\ref{fig:ebm}b). Most pair a price characteristic or a narrative exposure with the share of a firm's news in one topic, a feature that is imputed noise for 40\% to 50\% of firms. Deleting the pairs from the fitted model, with nothing re-estimated, raises the Sharpe ratio from $-0.23$ to 0.00, from $-0.52$ to 0.23 and from $-1.08$ to $-0.41$ (Figure~\ref{fig:ebm}c).

\subsection{A Longer Shock Window}
\label{sec:results:window}

\begin{figure}[tbp]
\centering
\includegraphics[width=\linewidth]{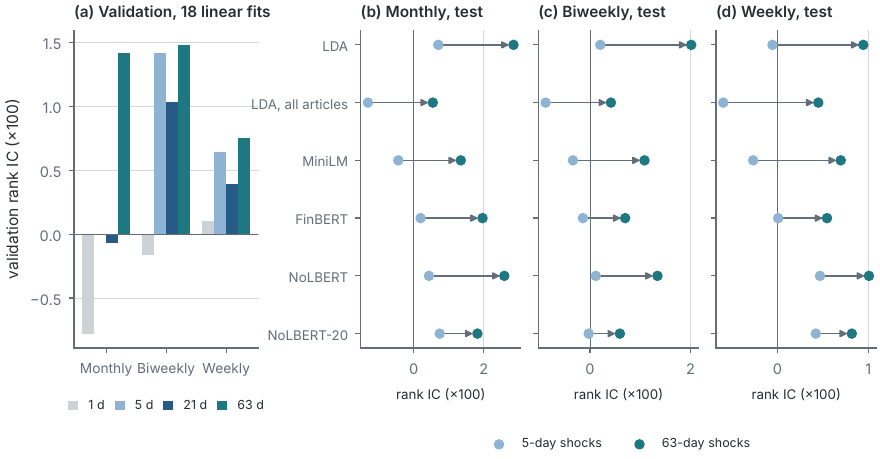}
\caption{The shock window. (a) Validation IC of the narrative exposures for windows of 1, 5, 21 and 63 trading days, averaged over OLS, ridge and lasso and the six readers. (b--d) Test IC of each reader's narrative, text and price-and-text experiments with 5-day and 63-day shocks.}
\label{fig:shockwindow}
\end{figure}

One choice made on validation does carry over. The rule of Section~\ref{sec:methods:attention} picked a 63-day window for the attention shocks at every frequency, although its validation advantage was small (Figure~\ref{fig:shockwindow}a). In the test period the longer window raises the IC of every reader at every frequency (Figure~\ref{fig:shockwindow}b--d). Over the 46 to 48 pairs of experiments that differ only in the window, the IC rises by 1.28, 0.87 and 0.49 ($t=1.8$, 2.0 and 1.7), the net long-short return by 5.8, 4.7 and 3.9 percentage points a year, and the Sharpe ratio improves in 91\%, 80\% and 85\% of pairs (Table~\ref{tab:improvements}). A 5-day shock records the news of one week. A 63-day shock records which topics are gaining attention relative to the past quarter, and stock exposures to those shifts forecast returns better. The gain is also stable across the two halves of the test period (Section~\ref{sec:results:robust}).

\subsection{Topic Readers}
\label{sec:results:readers}

\begin{figure}[tbp]
\centering
\includegraphics[width=\linewidth]{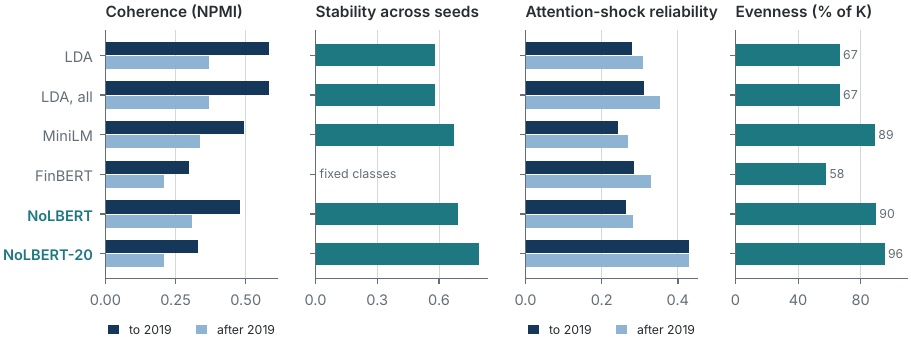}
\caption{The six readers. Coherence: mean NPMI \citep{roder2015coherence} of each topic's ten top words, on text up to 2019 and after. Stability: share of articles that keep their topic when the reader is refitted with another seed. Reliability: split-half reliability of the daily 5-day attention shocks. Evenness: effective number of topics as a share of $K$.}
\label{fig:topicquality}
\end{figure}

Figure~\ref{fig:topicquality} describes the readers themselves. When top words are chosen by the same rule for every reader, LDA has the most coherent topics (NPMI 0.58, against 0.48 for NoLBERT and 0.49 for MiniLM). The NoLBERT readers have the most stable topics, with 69\% and 80\% of articles keeping their topic when the model is refitted with another seed, against 58\% for LDA, and the most even topic sizes. Fewer topics give more reliable shocks: the split-half reliability of NoLBERT-20 is 0.47, against 0.28 for NoLBERT with 60 topics. On text after 2019 coherence falls by 29\% to 37\% for every reader, since none is refitted.

These differences matter less for forecasting than the shock window does. With 63-day shocks every reader has a positive average IC at every frequency, and LDA and NoLBERT lead (2.85 and 2.59 monthly, 2.01 and 1.34 biweekly, 0.94 and 1.01 weekly; Table~\ref{tab:readers}). NoLBERT is the only reader among the best three in every column of that table. Better shock reliability does not translate into better forecasts: LDA applied to all articles has more reliable shocks than LDA on the daily sample and forecasts worst of all six, and NoLBERT-20 does not beat NoLBERT. None of the pairwise differences between readers on the text-only sets is significant.

\subsection{Combining Readers and Models}
\label{sec:results:portfolios}

Single picks are unreliable and the 63-day window helps every reader, so we average all readers and models built with that window. Table~\ref{tab:main-M} reports the monthly results next to the experiments picked within each family, and Tables~\ref{tab:main-2W} and~\ref{tab:main-W} do the same for the other frequencies.

\begin{figure}[tbp]
\centering
\includegraphics[width=\linewidth]{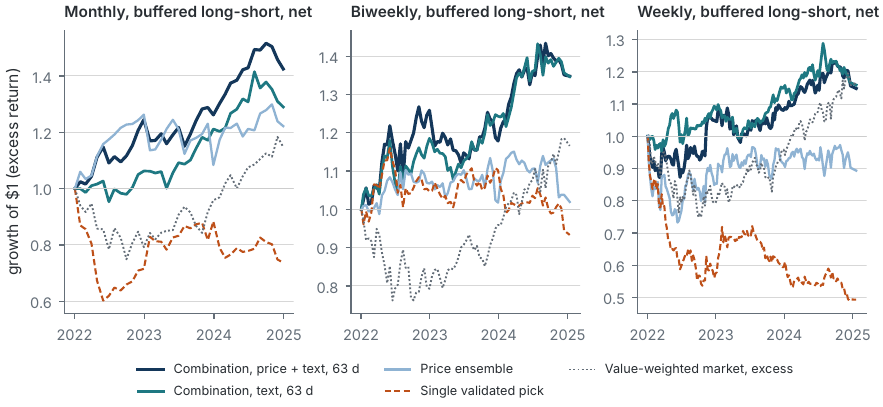}
\caption{Growth of \$1 in the buffered long-short portfolios, net of costs: the combinations across readers with 63-day shocks, the price ensemble and the overall pick, with the excess return of the value-weighted market for reference.}
\label{fig:cumret}
\end{figure}

Monthly, the combination on price and text with 63-day shocks has the largest IC of all (3.9, $t=2.91$) and the best Sharpe ratios, 1.12 for the decile portfolio and 1.19 buffered. The buffered portfolio earns 12.4\% a year net at a volatility of 10.4\%, with a maximum drawdown of 8\% and the lowest turnover in the table. On text alone the same combination has Sharpe ratios of 0.86 and 0.84. With 5-day shocks the two combinations drop to 0.39 and $-0.45$. The price ensemble has an IC of 3.3 and Sharpe ratios of 0.66 and 0.53. Figure~\ref{fig:cumret} shows how these portfolios grew.

The picture is similar biweekly, with Sharpe ratios of 0.69 and 0.77 for price and text and 0.63 and 0.86 for text alone, against 0.10 for the price ensemble. Weekly, costs take most of the return. The price-and-text combination has a gross Sharpe ratio of 1.02 and a net one of 0.30 (0.39 buffered), and the best weekly portfolios are single experiments, such as IPCA on price and NoLBERT-20 text (0.44 and 0.59). Both 63-day combinations are positive under both equal-weighted rules at every frequency. The overall pick is negative under both at every frequency.

\begin{table}[tbp]
\centering
\scriptsize
\caption{Monthly rebalancing: test-period forecasts and portfolios (36 months).}
\label{tab:main-M}
\setlength{\tabcolsep}{2.6pt}
\renewcommand{\arraystretch}{1.15}
\begin{tabular}{@{}L{3.05cm}L{4.15cm}rrrrrrrrrrr@{}}
\maintablehead
\multicolumn{13}{l}{\textit{Experiments selected within each family}}\\
Overall $\cdot$ Price\,+\,text $\cdot$ Every set $\cdot$ LDA & Price + text, LDA $\cdot$ \textit{EBM} & $+$1.5 & 0.53 & $-$0.23 & $-$0.37 & 0.06 & 0.05 & $-$7.9 & 21.2 & $-$40 & 80 & $-$0.26 \\
Price & Price $\cdot$ \textit{OLS} & $-$0.0 & $-$0.02 & $-$0.24 & $-$0.21 & 0.23 & 0.42 & $-$3.6 & 17.0 & $-$28 & 117 & $-$0.80 \\
Text $\cdot$ NoLBERT & Exposures, NoLBERT $\cdot$ \textit{IPCA} & $-$3.5 & $-$1.48 & $-$0.62 & $-$0.82 & $-$0.36 & $-$0.36 & $-$13.5 & 16.4 & $-$40 & \topC{63} & $-$1.55 \\
Tone/emb. & Price + all news $\cdot$ \textit{IPCA} & $+$2.5 & 1.64 & \topB{0.94} & \topB{1.04} & 0.26 & 0.32 & \topB{11.4} & 10.9 & $-$13 & 75 & \topC{1.54} \\
63\,d shocks & Text, LDA, 63 d $\cdot$ \textit{EBM} & $+$0.3 & 0.12 & $-$0.18 & $-$0.29 & \topA{0.63} & 0.45 & $-$6.0 & 20.5 & $-$34 & 100 & $-$0.03 \\
Ranked targets & Price + text, NoLBERT $\cdot$ \textit{IPCA, ranked} & $+$1.0 & 1.07 & $-$0.02 & 0.38 & $-$0.00 & \topC{0.46} & 3.1 & \topA{8.2} & $-$11 & 66 & 0.62 \\
Ensembles & Exposures, NoLBERT $\cdot$ \textit{Ensemble} & $+$0.1 & 0.06 & 0.04 & 0.00 & $-$0.51 & $-$0.69 & 0.0 & 11.1 & $-$20 & 68 & 0.32 \\
LDA (all articles) & Text, LDA (all articles) $\cdot$ \textit{EBM} & $-$0.7 & $-$0.33 & $-$0.62 & $-$0.61 & $-$0.46 & $-$0.43 & $-$10.7 & 17.5 & $-$42 & 88 & $-$0.82 \\
MiniLM & Price + text, MiniLM $\cdot$ \textit{EBM} & $-$6.8 & $-$2.71 & $-$1.76 & $-$1.71 & $-$1.02 & $-$0.63 & $-$33.6 & 19.6 & $-$68 & 103 & $-$3.79 \\
FinBERT & Text, FinBERT $\cdot$ \textit{IPCA} & $-$2.6 & $-$1.79 & $-$1.05 & $-$0.96 & $-$0.34 & $-$0.24 & $-$10.5 & 11.0 & $-$29 & 71 & $-$1.93 \\
NoLBERT-20 & Text, NoLBERT-20 $\cdot$ \textit{EBM} & $+$1.4 & 1.18 & 0.31 & 0.57 & \topC{0.38} & \topA{0.51} & 5.9 & \topB{10.3} & \topC{$-$10} & 79 & 1.25 \\
\addlinespace[2pt]\midrule
\multicolumn{13}{l}{\textit{Equal-weight combinations}}\\
Price ensemble & Price $\cdot$ \textit{Ensemble} & \topB{$+$3.3} & \topB{2.41} & 0.66 & 0.53 & \topB{0.40} & \topB{0.46} & 7.7 & 14.5 & $-$14 & 95 & 0.63 \\
All readers, 5 d & Price + text, all readers, 5 d $\cdot$ \textit{Ensemble} & $+$2.5 & 1.40 & 0.39 & 0.34 & $-$0.46 & $-$0.35 & 4.6 & 13.7 & $-$18 & 66 & 0.59 \\
All readers, 63 d & Price + text, all readers, 63 d $\cdot$ \textit{Ensemble} & \topA{$+$3.9} & \topA{2.91} & \topA{1.12} & \topA{1.19} & 0.03 & 0.12 & \topA{12.4} & \topC{10.4} & \topA{$-$8} & \topA{60} & \topA{2.21} \\
All readers, text, 5 d & Text, all readers, 5 d $\cdot$ \textit{Ensemble} & $-$1.0 & $-$0.82 & $-$0.45 & $-$0.48 & $-$0.23 & $-$0.20 & $-$5.8 & 12.3 & $-$20 & 65 & $-$0.70 \\
All readers, text, 63 d & Text, all readers, 63 d $\cdot$ \textit{Ensemble} & $+$2.5 & 1.90 & \topC{0.86} & \topC{0.84} & 0.32 & 0.30 & \topC{9.1} & 10.8 & \topB{$-$9} & \topB{61} & 1.48 \\
Best ten & Ten experiments $\cdot$ \textit{Best ten} & $+$2.4 & 1.58 & 0.33 & 0.57 & $-$0.07 & $-$0.13 & 6.3 & 10.9 & $-$12 & 77 & \topB{1.58} \\
Price ens. + tone & Price + tone $\cdot$ \textit{Combo} & \topC{$+$3.0} & \topC{2.37} & 0.69 & 0.70 & 0.34 & 0.41 & 8.5 & 12.2 & $-$11 & 119 & 1.04 \\
 
\bottomrule
\end{tabular}
\maintablenotes
\end{table}

\subsection{Portfolio Construction}
\label{sec:results:rules}

Figure~\ref{fig:rulesresults} summarizes every base-model experiment under each rule. The buffer helps at every frequency. The buffered long-short has the highest median Sharpe ratio (0.23, 0.10 and $-0.23$, against 0.16, $-0.03$ and $-0.34$ for the decile rule) with about a third less turnover. In matched pairs it adds 0.4, 1.1 and 1.4 percentage points a year (Table~\ref{tab:improvements}). Weighting the legs by market value lowers the median Sharpe ratio, because it puts most of each leg in the largest stocks, where the forecasts are weakest. The monthly price-and-text combination falls from 1.12 to 0.03 with market weights. Mean-variance weights, which rely on a noisy covariance estimate, do worst. The long-only portfolios and the tilt have median information ratios near or below zero. Table~\ref{tab:rules} gives the numbers.

Table~\ref{tab:improvements} collects the four choices we can test in matched pairs. The 63-day window and the ranked target improve both the forecast and the portfolio at every frequency; the ranked target is significant biweekly ($t=2.5$). The eight-model ensemble beats the average of its members in IC and has a higher Sharpe ratio in all six price-and-text sets monthly and biweekly. The buffer leaves the forecast unchanged and cuts turnover from 102\% to 73\% a month.

\begin{table}[tbp]
\centering
\footnotesize
\caption{Paired comparisons. Each pair of experiments differs only in the stated choice. $\Delta$IC: mean difference in IC; $\Delta$Ret.: mean difference in net long-short return (percentage points a year); $t$: Newey--West $t$ statistics; SR: median Sharpe ratio without and with the change; Impr.: share of pairs whose Sharpe ratio improves (\%); Turnover: median turnover per period (\%).}
\label{tab:improvements}
\setlength{\tabcolsep}{3.4pt}
\begin{tabular}{@{}L{3.6cm}lr rr rr rr r rr@{}}
\toprule
 & & & \multicolumn{2}{c}{Forecast} & \multicolumn{2}{c}{Return} & \multicolumn{2}{c}{SR} & & \multicolumn{2}{c}{Turnover} \\
\cmidrule(lr){4-5}\cmidrule(lr){6-7}\cmidrule(lr){8-9}\cmidrule(l){11-12}
Comparison & Frequency & Pairs & $\Delta$IC & $t$ & $\Delta$Ret. & $t$ & Base & New & Impr. & Base & New \\
\midrule
63-day against 5-day shocks & Monthly & 46 & \topB{$+$1.28} & 1.8 & \topB{$+$5.8} & \topA{2.2} & 0.24 & \topA{0.62} & \topC{91} & 106 & 102 \\
 & Biweekly & 46 & $+$0.87 & \topB{2.0} & \topC{$+$4.7} & \topB{2.0} & 0.05 & 0.34 & 80 & 81 & 80 \\
 & Weekly & 48 & $+$0.49 & 1.7 & $+$3.9 & 1.6 & $-$0.24 & 0.00 & 85 & 89 & 88 \\
\addlinespace[3pt]
Ranked against raw target & Monthly & 32 & $+$0.26 & 0.5 & $-$1.0 & $-$0.4 & 0.25 & \topC{0.37} & 59 & 102 & 101 \\
 & Biweekly & 34 & \topC{$+$0.95} & \topA{2.5} & $+$3.4 & 1.1 & 0.06 & 0.30 & 79 & 76 & 76 \\
 & Weekly & 36 & \topA{$+$1.57} & \topC{1.8} & \topA{$+$9.3} & 1.1 & $-$0.23 & 0.34 & 81 & 87 & 80 \\
\addlinespace[3pt]
Ensemble against its members & Monthly & 6 & $+$0.58 & 1.2 & $+$2.6 & 1.0 & 0.22 & \topB{0.55} & \topA{100} & 102 & 108 \\
 & Biweekly & 6 & $+$0.27 & 0.8 & $+$1.6 & 0.5 & 0.01 & 0.16 & \topB{100} & 77 & 83 \\
 & Weekly & 6 & $+$0.34 & 1.4 & $+$1.4 & 0.6 & $-$0.24 & $-$0.19 & 67 & 89 & 101 \\
\addlinespace[3pt]
Buffered against decile & Monthly & 48 & -- & -- & $+$0.4 & 1.7 & 0.24 & 0.26 & 65 & 102 & \topC{73} \\
 & Biweekly & 48 & -- & -- & $+$1.1 & \topC{1.8} & 0.05 & 0.13 & 89 & 77 & \topA{51} \\
 & Weekly & 48 & -- & -- & $+$1.4 & 1.5 & $-$0.24 & $-$0.20 & 73 & 89 & \topB{58} \\
 
\bottomrule
\end{tabular}
\tablenotes{\shadekey; base columns are not shaded and lower turnover is better.}
\end{table}

\subsection{Statistical Significance}
\label{sec:results:significance}

Table~\ref{tab:significance} tests the two 63-day combinations with their forecasts fixed, so that only the realized returns are random. Monthly, the price-and-text combination beats zero at the 5\% level ($t=2.36$, $p=0.045$; buffered $t=2.42$, $p=0.040$). It does not beat the price ensemble ($p=0.46$) or the same combination with 5-day shocks ($p=0.15$). On text alone the 63-day window does beat the 5-day window, monthly ($p=0.028$, buffered 0.025) and biweekly with the buffer ($p=0.048$). No deflated Sharpe ratio is close to 0.95. With about 600 experiments per frequency, the best of as many strategies without skill would be expected to reach a Sharpe ratio of 2.2 to 2.7.

Figure~\ref{fig:search} asks the same question of the best experiment overall. In the main design the best Sharpe ratio is below the no-skill benchmark for 24 of the 27 combinations of rule and frequency, and Hansen's test rejects once. When the 63-day sets are included, the reality check rejects twice and Hansen's test four times, all monthly or biweekly, and the winning portfolios all use the 63-day window. The best is the monthly decile portfolio of XGBoost on LDA exposures, with a Sharpe ratio of 2.15 against an expected best of 1.37 (reality check $p=0.016$). No experiment survives a 10\% false-discovery threshold. Exact values are in Table~\ref{tab:search}.

\begin{table}[tbp]
\centering
\footnotesize
\caption{Significance of the 63-day combinations across readers. SR: net Sharpe ratio; $t$, $p$: Newey--West test of the mean return with fixed-$b$ $p$-value; DSR: deflated Sharpe ratio. Each comparison gives the Sharpe ratio difference $\Delta$ and the $p$-value of the Ledoit--Wolf test.}
\label{tab:significance}
\setlength{\tabcolsep}{3.6pt}
\begin{tabular}{@{}llrrrr rr rr rr@{}}
\toprule
 & & \multicolumn{4}{c}{Against zero} & \multicolumn{2}{c}{vs.\ price ensemble} & \multicolumn{2}{c}{vs.\ 5-day shocks} & \multicolumn{2}{c}{vs.\ overall pick} \\
\cmidrule(lr){3-6}\cmidrule(lr){7-8}\cmidrule(lr){9-10}\cmidrule(l){11-12}
 & Combination, rule & SR & $t$ & $p$ & DSR & $\Delta$ & $p$ & $\Delta$ & $p$ & $\Delta$ & $p$ \\
\midrule
Monthly & Price + text, decile & \topB{1.12} & \topB{2.36} & \topB{0.045} & \topB{0.04} & $+$0.46 & 0.463 & $+$0.74 & 0.149 & $+$1.35 & 0.196 \\
 & Price + text, buffered & \topA{1.19} & \topA{2.42} & \topA{0.040} & \topA{0.06} & \topC{$+$0.66} & \topB{0.267} & $+$0.85 & 0.119 & \topA{$+$1.56} & 0.146 \\
 & Text only, decile & \topC{0.86} & 1.60 & 0.157 & 0.01 & $+$0.20 & 0.836 & \topB{$+$1.31} & \topB{0.028} & $+$1.09 & 0.182 \\
 & Text only, buffered & 0.84 & 1.77 & 0.122 & \topC{0.01} & $+$0.31 & 0.722 & \topA{$+$1.32} & \topA{0.025} & $+$1.21 & 0.129 \\
\addlinespace[3pt]
Biweekly & Price + text, decile & 0.69 & 1.41 & 0.182 & 0.00 & $+$0.59 & 0.266 & $+$0.46 & 0.152 & $+$0.96 & 0.101 \\
 & Price + text, buffered & 0.77 & 1.58 & 0.136 & 0.01 & \topB{$+$0.66} & \topA{0.211} & $+$0.40 & 0.236 & $+$0.90 & 0.147 \\
 & Text only, decile & 0.63 & 1.26 & 0.230 & 0.00 & $+$0.53 & 0.428 & $+$1.10 & 0.088 & $+$0.89 & 0.098 \\
 & Text only, buffered & 0.86 & \topC{1.84} & \topC{0.085} & 0.01 & \topA{$+$0.75} & 0.258 & \topC{$+$1.24} & \topC{0.048} & $+$0.99 & \topC{0.063} \\
\addlinespace[3pt]
Weekly & Price + text, decile & 0.30 & 0.56 & 0.591 & 0.00 & $+$0.34 & 0.408 & $+$0.29 & 0.422 & \topC{$+$1.37} & \topB{0.040} \\
 & Price + text, buffered & 0.39 & 0.77 & 0.459 & 0.00 & $+$0.46 & \topC{0.239} & $+$0.48 & 0.188 & $+$1.37 & \topA{0.024} \\
 & Text only, decile & 0.24 & 0.43 & 0.676 & 0.00 & $+$0.28 & 0.748 & $+$0.68 & 0.311 & $+$1.31 & 0.153 \\
 & Text only, buffered & 0.50 & 0.94 & 0.364 & 0.00 & $+$0.57 & 0.489 & $+$1.07 & 0.119 & \topB{$+$1.48} & 0.092 \\
 
\bottomrule
\end{tabular}
\tablenotes{\shadekey; a $p$-value is shaded with its $t$ statistic. Table~\ref{tab:significance-full} adds the bootstrap test and further benchmarks.}
\end{table}

\subsection{Robustness}
\label{sec:results:robust}

\begin{figure}[tbp]
\centering
\begin{subfigure}[t]{\linewidth}
\centering
\includegraphics[width=\linewidth]{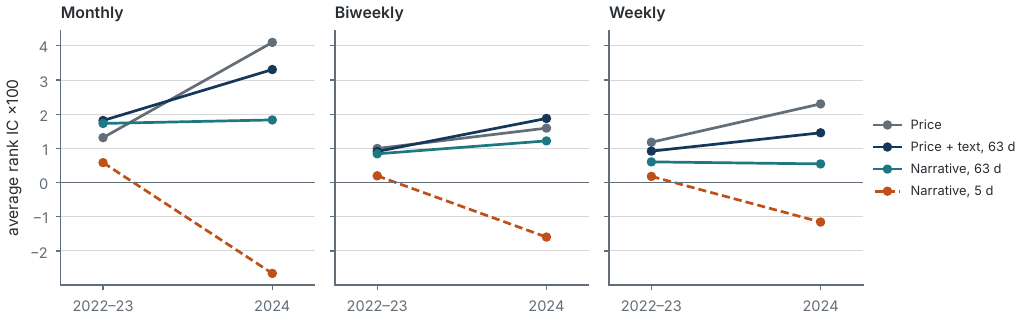}
\caption{Average test IC by kind of feature set, 2022--2023 and 2024.}
\end{subfigure}

\vspace{0.5em}
\begin{subfigure}[t]{\linewidth}
\centering
\includegraphics[width=0.92\linewidth]{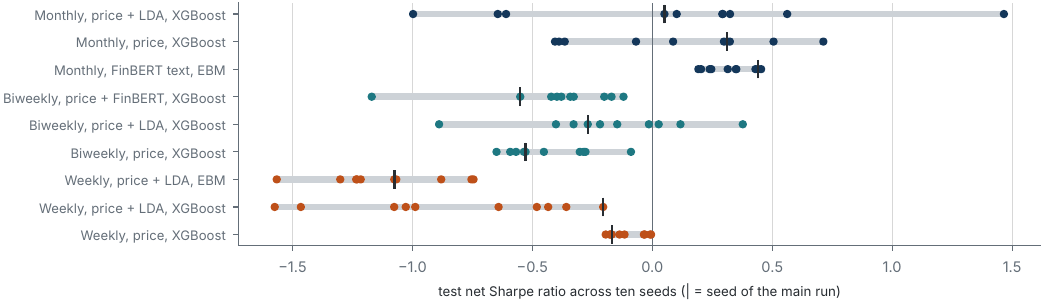}
\caption{Test Sharpe ratio of nine experiments rerun with ten seeds; the tick marks the seed of the main run.}
\end{subfigure}
\caption{Robustness to the period and to the seed.}
\label{fig:robust}
\end{figure}

The 2024 news comes from a second collection with fewer tickers and no headlines. Figure~\ref{fig:robust}a splits the test period. With 5-day shocks the IC of the narrative exposures turns from slightly positive in 2022--2023 to $-2.7$, $-1.6$ and $-1.2$ in 2024. With 63-day shocks it hardly changes (1.7 and 1.8 monthly). Price is stronger in 2024. None of the differences between the halves is significant.

Each experiment in the main tables is a single run. Figure~\ref{fig:robust}b repeats nine of them with ten seeds, tuning included. For a tree model on many features the seed alone moves the Sharpe ratio by as much as 2.5, from $-1.00$ to 1.47 for monthly XGBoost on price and LDA text, which is as large as the sampling error of three years of returns. This is a second reason to prefer combinations over single experiments.

\section{Discussion and Limitations}
\label{sec:discussion}

\paragraph{What a point-in-time encoder costs} Adapting NoLBERT took 150 minutes on one consumer GPU and no labels. The result is a reader whose pretraining text ends in 1995, whose topics are the most stable of the six, and whose forecasts are close to those of LDA and higher on average than those of MiniLM, an encoder trained on later text. We read this as evidence that the point-in-time restriction is affordable here. It is not evidence that it improves forecasts, and with three years of test data a moderate difference between readers in either direction would go undetected. The comparison with MiniLM also mixes two changes, the older pretraining text and our adaptation, which an adapted MiniLM run under the same recipe would separate.

\paragraph{Measuring attention} The window over which attention surprises are measured mattered more than the reader that measured them. A shock against the past quarter picks out topics that are gaining a lasting share of the news, and stock exposures to these slower shifts forecast returns in both halves of the test period, including the 2024 change of news source. The window was chosen on the validation period by a rule set in advance, but its advantage there was small, and we tried only four windows. The size of the test-period gain should therefore be read with the multiple-testing results in mind.

\paragraph{Choosing versus combining} A short validation period combined with flexible models produced the worst outcome of the study: the experiment with the best validation IC lost money at every frequency. The EBM's decomposition located the problem in pairwise terms built on a feature that is mostly imputed. Averaging over readers and models gave steadier results, as \citet{rapach2010outofsample} find for forecasts of the equity premium, and trading less through the buffer raised net returns in most experiments.

\paragraph{Limitations} The first limit is power. The test period has 36 months, 78 fortnights and 156 weeks, and the standard error of an annual Sharpe ratio is about 0.58 at every frequency, so only a true Sharpe ratio of about 1.7 or more would be detected four times out of five. Differences between readers, and between the combinations and the price ensemble, are not significant. The second is the search. Each frequency involves about 600 experiments and nine portfolio rules, no experiment passes a false-discovery threshold, and the deflated Sharpe ratios are small. The combinations across readers were formed after the main experiments had been run, so their $p$-values do not account for that choice.

The text models have their own limits. NoLBERT was adapted only on articles published up to the end of 2019, but MiniLM and the FinBERT classifier were trained on text from inside our sample. No reader's topics are refitted after 2019, and their coherence on later text is about a third lower. The news source was collected at the end of the sample: it lacks companies acquired during the sample and keeps only the recent articles of heavily covered firms, so firm-level news is imputed for 40\% to 50\% of stocks, and the 2024 articles come from a second collection with fewer tickers and no headlines. Finally, trading costs are a flat 10 basis points per dollar traded, with no borrowing costs for short positions and no price impact, the universe is limited to the 1,000 largest stocks, and each experiment is run with a single seed.

\section{Conclusion}
\label{sec:conclusion}

We built news topics from an encoder that has read nothing published after 1995 and tested them against readers trained on recent text, in a design where every model choice precedes the test period. The point-in-time encoder gave stable topics and forecasts comparable to the best alternative. Measuring attention shocks over a quarter improved forecasts for every reader, and combining readers and models produced the strongest portfolios, with a monthly net Sharpe ratio of 1.12. That result is significant against zero before adjusting for the search, but it does not beat price-based forecasts and its deflated Sharpe ratio is small. A longer test period, a modern encoder adapted under the same recipe, and topics refitted over time would sharpen these comparisons.

\section{Data and Code Availability}
\label{sec:availability}

The article text comes from FNSPID \citep{dong2024fnspid} and the article list from FinMultiTime \citep{xu2025finmultitime}, both public on Hugging Face. Stock data come from CRSP through WRDS \citep{crspstockdatabase}. They are licensed and not redistributed, but the repository includes the script that rebuilds the universe and return file for users with WRDS access. Factors and the risk-free rate are from the Kenneth R.\ French Data Library \citep{frenchlibrary}. Code, derived topic outputs, forecasts and the scripts behind every table and figure are in the project repository \todo{link}.

\bibliographystyle{plainnat}
\bibliography{references}

\clearpage
\appendix
\setcounter{table}{0}
\renewcommand{\thetable}{A\arabic{table}}
\setcounter{figure}{0}
\renewcommand{\thefigure}{A\arabic{figure}}

\section{Additional Figures and Tables}
\label{app:tables}

\begin{figure}[!htbp]
\centering
\includegraphics[width=\linewidth]{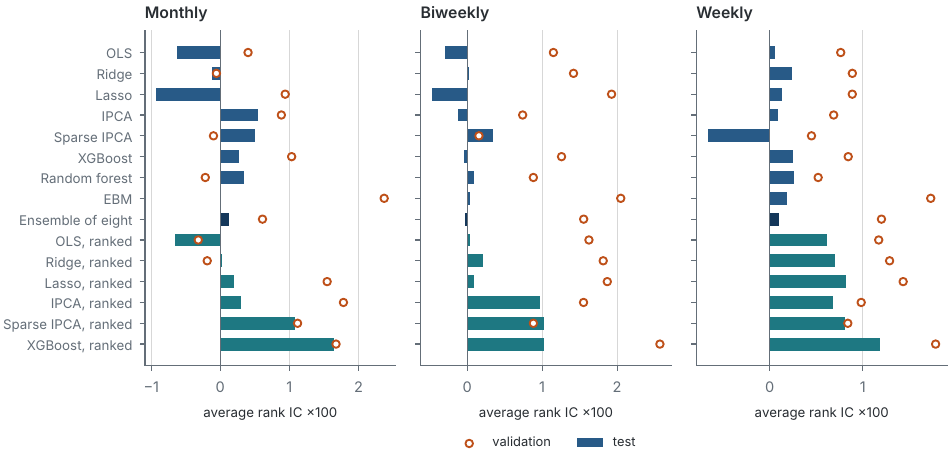}
\caption{Average IC of each model over the 26 feature sets of the main design, in the test period (bars; teal for models fitted on ranked returns) and the validation period (circles).}
\label{fig:models}
\end{figure}

\begin{figure}[!htbp]
\centering
\includegraphics[width=\linewidth]{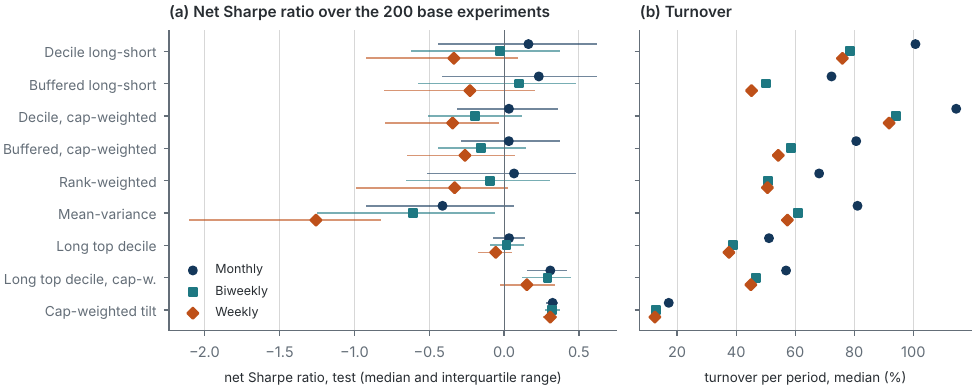}
\caption{Portfolio rules over the base-model experiments. (a) Median and interquartile range of the Sharpe ratio (information ratio against the universe for the long-only rules and the tilt). (b) Median turnover per period.}
\label{fig:rulesresults}
\end{figure}

\begin{figure}[!htbp]
\centering
\includegraphics[width=\linewidth]{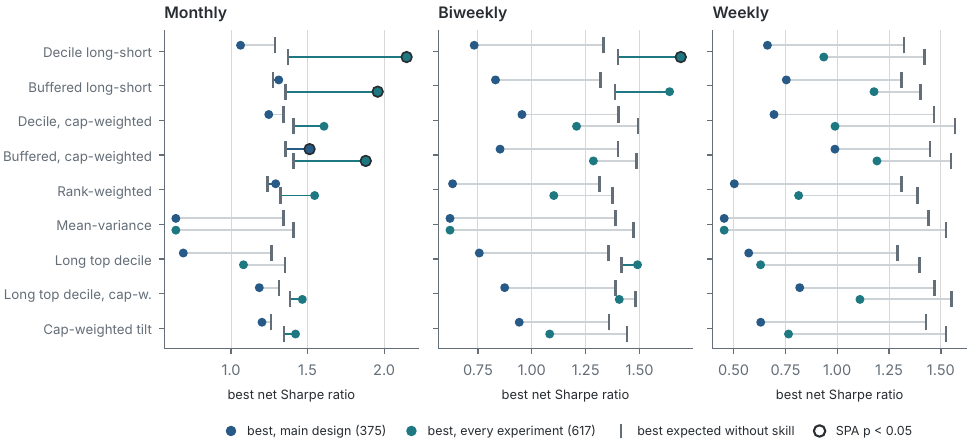}
\caption{The best of all experiments by rule and frequency, in the main design (375 experiments) and with every experiment (617), against the best Sharpe ratio expected without skill. A line is coloured when the best beats that benchmark.}
\label{fig:search}
\end{figure}

Figures~\ref{fig:models}--\ref{fig:search} support Section~\ref{sec:results}. The tables give the values behind its figures and the results of Table~\ref{tab:main-M} for the biweekly and weekly frequencies.

\begin{table}[!ht]
\centering
\footnotesize
\caption{Test-period forecasts by kind of feature set. IC: average IC of the base-model experiments; $t$: $t$ statistic of the group's average IC per period; Pos.: share of experiments with a positive IC (\%); SR: median Sharpe ratio of their decile long-short portfolios.}
\label{tab:kinds}
\setlength{\tabcolsep}{3.4pt}
\begin{tabular}{@{}lr rrrr rrrr rrrr@{}}
\toprule
 & & \multicolumn{4}{c}{Monthly} & \multicolumn{4}{c}{Biweekly} & \multicolumn{4}{c}{Weekly} \\
\cmidrule(lr){3-6}\cmidrule(lr){7-10}\cmidrule(l){11-14}
Kind of feature set & $n$ & IC & $t$ & Pos. & SR & IC & $t$ & Pos. & SR & IC & $t$ & Pos. & SR \\
\midrule
Price & 8 & \topC{$+$2.2} & \topC{2.4} & 75 & 0.41 & \topC{$+$1.2} & 1.7 & \topA{100} & $-$0.07 & \topA{$+$1.6} & 1.5 & \topC{88} & \topA{0.09} \\
Narrative exposures, 5 d & 48 & $-$0.5 & $-$0.9 & 40 & 0.03 & $-$0.4 & $-$0.7 & 31 & $-$0.20 & $-$0.3 & $-$0.7 & 40 & $-$0.46 \\
Narrative exposures, 63 d & 48 & $+$1.8 & 2.1 & \topA{90} & \topA{0.84} & $+$1.0 & 1.5 & 85 & \topA{0.39} & $+$0.6 & 1.3 & 83 & \topB{0.03} \\
Firm news & 48 & $-$0.2 & $-$0.4 & 31 & $-$0.82 & $+$0.1 & 0.2 & 50 & $-$1.30 & $-$0.0 & $-$0.1 & 42 & $-$1.53 \\
Text, 5 d & 48 & $-$0.5 & $-$0.8 & 35 & 0.00 & $-$0.5 & $-$0.8 & 27 & $-$0.20 & $-$0.4 & $-$1.0 & 27 & $-$0.57 \\
Text, 63 d & 48 & $+$1.5 & 1.9 & \topC{83} & \topB{0.64} & $+$0.9 & 1.5 & 85 & \topC{0.30} & $+$0.5 & 1.3 & 81 & $-$0.06 \\
Price + text, 5 d & 48 & $+$1.1 & 1.5 & 60 & 0.24 & $+$0.4 & 0.7 & 48 & 0.05 & $+$0.6 & 1.4 & 75 & $-$0.24 \\
Price + text, 63 d & 48 & \topB{$+$2.3} & \topB{3.1} & \topB{90} & \topC{0.63} & \topB{$+$1.2} & \topC{2.1} & \topC{94} & \topB{0.34} & \topC{$+$1.1} & \topC{2.1} & 88 & \topC{0.00} \\
Tone (FinBERT, LM) & 16 & $+$0.6 & 1.6 & 81 & 0.06 & $+$0.3 & 1.0 & 81 & $-$1.00 & $-$0.1 & $-$0.7 & 19 & $-$1.96 \\
Headline embeddings & 8 & $-$0.5 & $-$0.6 & 12 & $-$1.12 & $+$0.9 & \topB{2.8} & 88 & $-$0.16 & $+$0.2 & 1.1 & 62 & $-$2.40 \\
Tone + embeddings & 8 & $-$0.3 & $-$0.6 & 25 & $-$0.92 & $+$0.6 & 1.7 & 62 & $-$0.66 & $+$0.4 & \topA{3.0} & 88 & $-$2.11 \\
Price + news & 24 & \topA{$+$2.4} & \topA{3.7} & 83 & 0.54 & \topA{$+$1.5} & \topA{2.8} & \topB{100} & 0.19 & \topB{$+$1.5} & \topB{2.1} & \topB{92} & $-$0.01 \\
Placebo exposures & 8 & $-$0.0 & $-$0.0 & 38 & $-$0.10 & $+$0.6 & 0.8 & 62 & 0.16 & $+$0.6 & 1.2 & \topA{100} & $-$0.02 \\
 
\bottomrule
\end{tabular}
\tablenotes{\shadekey. $n$: experiments per frequency.}
\end{table}

\begin{table}[!ht]
\centering
\footnotesize
\caption{Test-period forecasts by reader: average IC of the base-model experiments on the reader's narrative, text and price-and-text sets with 5-day or 63-day shocks, and the median Sharpe ratio (SR) of their decile long-short portfolios.}
\label{tab:readers}
\setlength{\tabcolsep}{3.6pt}
\begin{tabular}{@{}l rrrr rrrr rrrr@{}}
\toprule
 & \multicolumn{4}{c}{Monthly} & \multicolumn{4}{c}{Biweekly} & \multicolumn{4}{c}{Weekly} \\
\cmidrule(lr){2-5}\cmidrule(lr){6-9}\cmidrule(l){10-13}
 & \multicolumn{2}{c}{IC} & \multicolumn{2}{c}{SR} & \multicolumn{2}{c}{IC} & \multicolumn{2}{c}{SR} & \multicolumn{2}{c}{IC} & \multicolumn{2}{c}{SR} \\
\cmidrule(lr){2-3}\cmidrule(lr){4-5}\cmidrule(lr){6-7}\cmidrule(lr){8-9}\cmidrule(lr){10-11}\cmidrule(l){12-13}
Reader & 5\,d & 63\,d & 5\,d & 63\,d & 5\,d & 63\,d & 5\,d & 63\,d & 5\,d & 63\,d & 5\,d & 63\,d \\
\midrule
LDA & \topB{$+$0.70} & \topA{$+$2.85} & 0.11 & \topA{1.20} & \topA{$+$0.20} & \topA{$+$2.01} & \topB{0.04} & \topA{0.81} & $-$0.06 & \topB{$+$0.94} & $-$0.36 & \topA{0.26} \\
LDA (all articles) & $-$1.30 & $+$0.55 & $-$0.85 & $-$0.13 & $-$0.88 & $+$0.41 & $-$0.79 & $-$0.13 & $-$0.60 & $+$0.45 & $-$0.94 & $-$0.26 \\
MiniLM & $-$0.44 & $+$1.34 & $-$0.02 & 0.62 & $-$0.34 & \topC{$+$1.08} & $-$0.45 & 0.35 & $-$0.27 & $+$0.70 & $-$0.64 & $-$0.05 \\
FinBERT & $+$0.20 & \topC{$+$1.97} & \topB{0.20} & \topC{0.80} & $-$0.14 & $+$0.70 & $-$0.21 & 0.29 & \topC{$+$0.01} & $+$0.55 & \topC{$-$0.29} & $-$0.13 \\
NoLBERT & \topC{$+$0.44} & \topB{$+$2.59} & \topA{0.23} & \topB{0.92} & \topB{$+$0.11} & \topB{$+$1.34} & \topC{$-$0.01} & \topB{0.48} & \topA{$+$0.47} & \topA{$+$1.01} & \topB{$-$0.19} & \topC{0.16} \\
NoLBERT-20 & \topA{$+$0.74} & $+$1.82 & \topC{0.18} & 0.65 & \topC{$-$0.03} & $+$0.59 & \topA{0.08} & \topC{0.37} & \topB{$+$0.42} & \topC{$+$0.82} & \topA{$-$0.05} & \topB{0.23} \\
 
\bottomrule
\end{tabular}
\tablenotes{\shadekey.}
\end{table}

\begin{table}[!ht]
\centering
\footnotesize
\caption{Forecasts by model, averaged over the 26 feature sets of the main design. Val.: validation IC; IC and $t$: test IC and its $t$ statistic; SR: median test Sharpe ratio of the decile long-short portfolios.}
\label{tab:models-results}
\setlength{\tabcolsep}{3.6pt}
\begin{tabular}{@{}l rrrr rrrr rrrr@{}}
\toprule
 & \multicolumn{4}{c}{Monthly} & \multicolumn{4}{c}{Biweekly} & \multicolumn{4}{c}{Weekly} \\
\cmidrule(lr){2-5}\cmidrule(lr){6-9}\cmidrule(l){10-13}
Model & Val. & IC & $t$ & SR & Val. & IC & $t$ & SR & Val. & IC & $t$ & SR \\
\midrule
OLS & $+$0.4 & $-$0.6 & $-$1.3 & $-$0.20 & $+$1.1 & $-$0.3 & $-$0.6 & $-$0.45 & $+$0.8 & $+$0.1 & 0.2 & $-$0.61 \\
Ridge & $-$0.1 & $-$0.1 & $-$0.2 & \topB{0.20} & $+$1.4 & $+$0.0 & 0.0 & $-$0.15 & $+$0.9 & $+$0.2 & 0.7 & $-$0.34 \\
Lasso & $+$0.9 & $-$0.9 & $-$1.3 & $-$0.45 & $+$1.9 & $-$0.5 & $-$0.9 & $-$0.31 & $+$0.9 & $+$0.1 & 0.3 & $-$0.41 \\
IPCA & $+$0.9 & \topC{$+$0.6} & 1.1 & $-$0.02 & $+$0.7 & $-$0.1 & $-$0.4 & $-$0.41 & $+$0.7 & $+$0.1 & 0.4 & $-$0.44 \\
Sparse IPCA & $-$0.1 & $+$0.5 & \topC{1.1} & $-$0.17 & $+$0.2 & $+$0.3 & 0.8 & \topC{0.06} & $+$0.4 & $-$0.7 & $-$1.1 & $-$0.55 \\
XGBoost & $+$1.0 & $+$0.3 & 0.7 & $-$0.28 & $+$1.3 & $-$0.0 & $-$0.1 & $-$0.44 & $+$0.8 & $+$0.3 & 1.1 & $-$0.81 \\
Random forest & $-$0.2 & $+$0.3 & 0.7 & \topC{0.13} & $+$0.9 & $+$0.1 & 0.3 & $-$0.19 & $+$0.5 & $+$0.3 & 1.5 & $-$0.75 \\
EBM & $+$2.4 & $+$0.0 & 0.0 & $-$0.37 & $+$2.0 & $+$0.0 & 0.1 & $-$0.14 & $+$1.7 & $+$0.2 & 0.5 & $-$0.42 \\
\addlinespace[2pt]
Ensemble & $+$0.6 & $+$0.1 & 0.3 & 0.02 & $+$1.6 & $-$0.0 & $-$0.1 & $-$0.32 & $+$1.2 & $+$0.1 & 0.3 & $-$0.42 \\
\addlinespace[2pt]
OLS, ranked & $-$0.3 & $-$0.7 & $-$1.2 & $-$0.20 & $+$1.6 & $+$0.0 & 0.1 & $-$0.44 & $+$1.2 & $+$0.6 & \topC{1.8} & $-$0.35 \\
Ridge, ranked & $-$0.2 & $+$0.0 & 0.0 & $-$0.01 & $+$1.8 & $+$0.2 & 0.4 & $-$0.09 & $+$1.3 & $+$0.7 & 1.7 & \topB{$-$0.05} \\
Lasso, ranked & $+$1.5 & $+$0.2 & 0.2 & 0.03 & $+$1.9 & $+$0.1 & 0.2 & $-$0.15 & $+$1.4 & \topB{$+$0.8} & \topB{1.9} & \topA{$-$0.02} \\
IPCA, ranked & $+$1.8 & $+$0.3 & 0.7 & $-$0.38 & $+$1.6 & \topC{$+$1.0} & \topA{1.7} & $-$0.01 & $+$1.0 & $+$0.7 & 1.7 & $-$0.19 \\
Sparse IPCA, ranked & $+$1.1 & \topB{$+$1.1} & \topB{1.6} & $-$0.18 & $+$0.9 & \topA{$+$1.0} & \topB{1.6} & \topA{0.11} & $+$0.8 & \topC{$+$0.8} & 1.5 & \topC{$-$0.09} \\
XGBoost, ranked & $+$1.7 & \topA{$+$1.6} & \topA{1.8} & \topA{0.22} & $+$2.6 & \topB{$+$1.0} & \topC{1.5} & \topB{0.09} & $+$1.8 & \topA{$+$1.2} & \topA{2.6} & $-$0.10 \\
 
\bottomrule
\end{tabular}
\tablenotes{\shadekey; validation columns are not shaded.}
\end{table}

\begin{table}[!ht]
\centering
\footnotesize
\caption{Portfolio rules over the base-model experiments. SR: median test Sharpe ratio; Pos.: share of experiments with a positive one (\%); Turn.: median turnover per period (\%).}
\label{tab:rules}
\setlength{\tabcolsep}{4.2pt}
\begin{tabular}{@{}l rrr rrr rrr@{}}
\toprule
 & \multicolumn{3}{c}{Monthly} & \multicolumn{3}{c}{Biweekly} & \multicolumn{3}{c}{Weekly} \\
\cmidrule(lr){2-4}\cmidrule(lr){5-7}\cmidrule(l){8-10}
Rule & SR & Pos. & Turn. & SR & Pos. & Turn. & SR & Pos. & Turn. \\
\midrule
Decile long-short & \topB{0.16} & \topA{57} & 101 & \topB{$-$0.03} & \topB{48} & 79 & $-$0.34 & \topB{30} & 76 \\
Buffered long-short & \topA{0.23} & \topB{57} & 72 & \topA{0.10} & \topA{54} & 50 & \topA{$-$0.23} & \topA{38} & 45 \\
Decile, cap-weighted & 0.03 & 50 & 114 & $-$0.20 & 31 & 94 & $-$0.34 & 22 & 92 \\
Buffered, cap-weighted & 0.03 & \topC{51} & 81 & $-$0.16 & 35 & 59 & \topB{$-$0.26} & \topC{30} & 54 \\
Rank-weighted & \topC{0.07} & 51 & 68 & $-$0.10 & 44 & 51 & $-$0.33 & 26 & 51 \\
Mean-variance & $-$0.41 & 28 & 81 & $-$0.61 & 21 & 61 & $-$1.26 & 2 & 57 \\
Long top decile$^\dagger$ & $-$0.10 & 42 & \topB{51} & $-$0.19 & 35 & \topB{39} & $-$0.40 & 18 & \topB{38} \\
Long top decile, cap-w.$^\dagger$ & 0.02 & 50 & \topC{57} & \topC{$-$0.04} & \topC{45} & \topC{47} & \topC{$-$0.28} & 29 & \topC{45} \\
Cap-weighted tilt$^\dagger$ & $-$0.14 & 37 & \topA{17} & $-$0.37 & 20 & \topA{13} & $-$0.44 & 16 & \topA{12} \\
 
\bottomrule
\end{tabular}
\tablenotes{\shadekey; lower turnover is better. $^\dagger$Information ratio against the universe.}
\end{table}

\begin{table}[!ht]
\centering
\scriptsize
\caption{The best of the search. Best: best test Sharpe ratio (information ratio for $^\dagger$); No skill: best expected without skill, from the bootstrap with every mean return set to zero; SPA: $p$-value of Hansen's test.}
\label{tab:search}
\setlength{\tabcolsep}{2.4pt}
\begin{tabular}{@{}l rrr rrr rrr rrr rrr rrr@{}}
\toprule
 & \multicolumn{6}{c}{Monthly} & \multicolumn{6}{c}{Biweekly} & \multicolumn{6}{c}{Weekly} \\
\cmidrule(lr){2-7}\cmidrule(lr){8-13}\cmidrule(l){14-19}
 & \multicolumn{3}{c}{Main (375)} & \multicolumn{3}{c}{All (617)} & \multicolumn{3}{c}{Main (375)} & \multicolumn{3}{c}{All (617)} & \multicolumn{3}{c}{Main (375)} & \multicolumn{3}{c}{All (617)} \\
\cmidrule(lr){2-4}\cmidrule(lr){5-7}\cmidrule(lr){8-10}\cmidrule(lr){11-13}\cmidrule(lr){14-16}\cmidrule(l){17-19}
Rule & Best & \makecell[c]{No\\skill} & SPA & Best & \makecell[c]{No\\skill} & SPA & Best & \makecell[c]{No\\skill} & SPA & Best & \makecell[c]{No\\skill} & SPA & Best & \makecell[c]{No\\skill} & SPA & Best & \makecell[c]{No\\skill} & SPA \\
\midrule
Decile long-short & 1.06 & 1.29 & 0.59 & \topA{2.15} & 1.37 & \topA{0.00} & 0.73 & 1.33 & 0.88 & \topA{1.69} & 1.40 & \topA{0.03} & 0.66 & 1.32 & \topB{0.93} & 0.93 & 1.42 & 0.81 \\
Buffered long-short & \topB{1.31} & 1.27 & \topC{0.16} & \topB{1.96} & 1.36 & \topB{0.01} & 0.83 & 1.32 & \topC{0.72} & \topB{1.64} & 1.39 & \topB{0.06} & \topC{0.75} & 1.31 & 0.96 & \topB{1.18} & 1.40 & \topA{0.56} \\
Decile, cap-weighted & 1.25 & 1.34 & \topB{0.16} & 1.61 & 1.41 & 0.21 & \topA{0.95} & 1.40 & \topB{0.69} & 1.21 & 1.49 & 0.70 & 0.69 & 1.47 & \topC{0.94} & 0.99 & 1.57 & 0.91 \\
Buffered, cap-weighted & \topA{1.51} & 1.36 & \topA{0.02} & \topC{1.88} & 1.41 & \topC{0.03} & 0.85 & 1.40 & 0.95 & 1.29 & 1.48 & 0.56 & \topA{0.99} & 1.45 & \topA{0.69} & \topA{1.19} & 1.55 & \topB{0.73} \\
Rank-weighted & \topC{1.29} & 1.24 & 0.29 & 1.55 & 1.32 & 0.13 & 0.63 & 1.31 & 0.96 & 1.10 & 1.37 & 0.53 & 0.50 & 1.31 & 0.99 & 0.81 & 1.39 & 0.94 \\
Mean-variance & 0.64 & 1.34 & 0.99 & 0.64 & 1.41 & 0.99 & 0.62 & 1.39 & 0.98 & 0.62 & 1.47 & 0.99 & 0.45 & 1.44 & 1.00 & 0.45 & 1.52 & 1.00 \\
Long top decile$^\dagger$ & 0.69 & 1.26 & 0.91 & 1.08 & 1.35 & 0.57 & 0.76 & 1.35 & 0.98 & \topC{1.49} & 1.42 & \topC{0.07} & 0.57 & 1.29 & 0.99 & 0.63 & 1.40 & 0.99 \\
Long top decile, cap-w.$^\dagger$ & 1.18 & 1.31 & 0.36 & 1.47 & 1.39 & 0.44 & \topC{0.87} & 1.39 & 0.94 & 1.41 & 1.48 & 0.50 & \topB{0.82} & 1.47 & 0.97 & \topC{1.11} & 1.55 & \topC{0.77} \\
Cap-weighted tilt$^\dagger$ & 1.20 & 1.26 & 0.30 & 1.42 & 1.35 & 0.16 & \topB{0.94} & 1.36 & \topA{0.62} & 1.08 & 1.44 & 0.72 & 0.63 & 1.43 & 0.98 & 0.76 & 1.52 & 0.99 \\
 
\bottomrule
\end{tabular}
\tablenotes{\shadekey; no-skill columns are not shaded and lower $p$-values are better.}
\end{table}

\begin{table}[!p]
\centering
\scriptsize
\caption{Biweekly rebalancing: test-period forecasts and portfolios (78 fortnights).}
\label{tab:main-2W}
\setlength{\tabcolsep}{2.6pt}
\renewcommand{\arraystretch}{1.15}
\begin{tabular}{@{}L{3.05cm}L{4.15cm}rrrrrrrrrrr@{}}
\maintablehead
\multicolumn{13}{l}{\textit{Experiments selected within each family}}\\
Overall $\cdot$ Text $\cdot$ FinBERT & Text, FinBERT $\cdot$ \textit{EBM} & $-$0.5 & $-$0.45 & $-$0.27 & $-$0.13 & $-$0.76 & $-$0.26 & $-$1.6 & 12.2 & $-$20 & 50 & $-$0.20 \\
Price & Price $\cdot$ \textit{IPCA} & $+$0.7 & 0.45 & 0.04 & $-$0.01 & $-$0.24 & $-$0.27 & $-$0.3 & 17.3 & $-$19 & 95 & $-$0.12 \\
Price\,+\,text $\cdot$ LDA & Price + text, LDA $\cdot$ \textit{EBM} & $-$1.2 & $-$0.64 & $-$0.52 & $-$0.63 & $-$0.76 & $-$0.95 & $-$14.3 & 22.9 & $-$47 & 77 & $-$1.19 \\
Tone/emb. & Price + all news $\cdot$ \textit{Lasso} & $+$0.8 & 0.74 & 0.07 & 0.21 & $-$0.41 & $-$0.39 & 2.5 & 12.2 & $-$15 & 45 & 0.55 \\
63\,d shocks $\cdot$ Every set & Text, MiniLM, 63 d $\cdot$ \textit{Sparse IPCA} & $+$0.1 & 0.10 & 0.19 & 0.39 & $-$0.14 & \topC{0.20} & 4.1 & \topC{10.6} & $-$18 & 42 & 0.79 \\
Ranked targets & Price + text, NoLBERT $\cdot$ \textit{XGBoost, ranked} & $+$1.2 & 1.02 & 0.18 & 0.21 & $-$0.46 & $-$0.28 & 2.6 & 12.3 & $-$21 & 66 & 0.52 \\
Ensembles & Text, NoLBERT $\cdot$ \textit{Ensemble} & $-$0.1 & $-$0.10 & $-$0.20 & 0.12 & $-$0.79 & $-$0.31 & 1.3 & 11.4 & $-$20 & 44 & 0.41 \\
LDA (all articles) & Exposures, LDA (all articles) $\cdot$ \textit{IPCA} & $-$0.1 & $-$0.14 & $-$0.21 & $-$0.32 & \topA{0.26} & \topB{0.34} & $-$3.6 & 11.3 & $-$23 & \topA{39} & $-$1.00 \\
MiniLM & Text, MiniLM $\cdot$ \textit{EBM} & $+$0.5 & 0.38 & $-$0.02 & 0.17 & $-$0.66 & $-$0.36 & 2.2 & 12.6 & $-$14 & 51 & 0.38 \\
NoLBERT & Price + text, NoLBERT $\cdot$ \textit{Ridge} & $-$0.1 & $-$0.14 & $-$0.03 & 0.14 & $-$0.16 & $-$0.18 & 1.4 & \topA{10.3} & $-$16 & 46 & 0.40 \\
NoLBERT-20 & Price + text, NoLBERT-20 $\cdot$ \textit{EBM} & $+$1.0 & 0.82 & 0.12 & 0.14 & $-$0.65 & $-$0.23 & 2.0 & 14.8 & $-$18 & 49 & 0.43 \\
\addlinespace[2pt]\midrule
\multicolumn{13}{l}{\textit{Equal-weight combinations}}\\
Price ensemble & Price $\cdot$ \textit{Ensemble} & \topB{$+$1.7} & \topB{1.66} & 0.10 & 0.11 & $-$0.26 & $-$0.29 & 1.4 & 12.2 & \topA{$-$11} & 97 & $-$0.11 \\
All readers, 5 d & Price + text, all readers, 5 d $\cdot$ \textit{Ensemble} & $+$0.8 & 0.65 & \topC{0.23} & 0.37 & $-$0.41 & $-$0.11 & \topC{5.2} & 14.0 & $-$14 & 46 & 0.75 \\
All readers, 63 d & Price + text, all readers, 63 d $\cdot$ \textit{Ensemble} & \topA{$+$2.4} & \topA{2.09} & \topA{0.69} & \topB{0.77} & $-$0.21 & $-$0.04 & \topA{11.0} & 14.2 & $-$12 & 45 & \topB{1.74} \\
All readers, text, 5 d & Text, all readers, 5 d $\cdot$ \textit{Ensemble} & $-$0.9 & $-$0.75 & $-$0.47 & $-$0.38 & $-$0.29 & $-$0.47 & $-$4.3 & 11.2 & $-$16 & \topC{41} & $-$0.43 \\
All readers, text, 63 d & Text, all readers, 63 d $\cdot$ \textit{Ensemble} & \topC{$+$1.6} & \topC{1.53} & \topB{0.63} & \topA{0.86} & \topC{0.07} & \topA{0.66} & \topB{10.7} & 12.5 & \topB{$-$11} & \topB{39} & \topA{2.07} \\
Best ten & Ten experiments $\cdot$ \textit{Best ten} & $-$0.2 & $-$0.14 & 0.12 & \topC{0.40} & $-$0.16 & $-$0.09 & 5.1 & 12.7 & \topC{$-$12} & 44 & \topC{1.03} \\
Price ens. + tone & Price + tone $\cdot$ \textit{Combo} & $+$1.2 & 1.29 & 0.00 & 0.02 & \topB{0.19} & $-$0.08 & 0.2 & \topB{10.6} & $-$15 & 112 & $-$0.26 \\
 
\bottomrule
\end{tabular}
\maintablenotes
\end{table}
\begin{table}[!p]
\centering
\scriptsize
\caption{Weekly rebalancing: test-period forecasts and portfolios (156 weeks).}
\label{tab:main-W}
\setlength{\tabcolsep}{2.6pt}
\renewcommand{\arraystretch}{1.15}
\begin{tabular}{@{}L{3.05cm}L{4.15cm}rrrrrrrrrrr@{}}
\maintablehead
\multicolumn{13}{l}{\textit{Experiments selected within each family}}\\
Overall $\cdot$ Price\,+\,text $\cdot$ Every set $\cdot$ Ranked targets $\cdot$ LDA & Price + text, LDA $\cdot$ \textit{EBM} & $-$1.3 & $-$0.91 & $-$1.08 & $-$0.98 & $-$1.01 & $-$0.88 & $-$21.2 & 21.6 & $-$51 & 88 & $-$1.87 \\
Price & Price $\cdot$ \textit{IPCA} & \topB{$+$3.0} & \topB{2.31} & \topB{0.43} & \topB{0.52} & $-$0.30 & $-$0.19 & \topB{9.5} & 18.3 & $-$15 & 106 & \topB{0.84} \\
Text $\cdot$ NoLBERT & Text, NoLBERT $\cdot$ \textit{EBM} & $+$0.3 & 0.35 & $-$0.08 & $-$0.04 & $-$0.16 & 0.02 & $-$0.6 & 14.0 & $-$17 & \topC{30} & 0.28 \\
Tone/emb. & Price + tone $\cdot$ \textit{IPCA} & \topC{$+$3.0} & \topC{2.30} & \topC{0.38} & 0.46 & $-$0.18 & $-$0.14 & \topC{8.4} & 18.4 & $-$17 & 105 & 0.75 \\
63\,d shocks & Price + text, LDA, 63 d $\cdot$ \textit{EBM} & $-$1.3 & $-$0.93 & $-$0.78 & $-$0.81 & $-$0.87 & $-$0.64 & $-$18.6 & 23.0 & $-$49 & 93 & $-$1.77 \\
Ensembles & Price + text, FinBERT $\cdot$ \textit{Ensemble} & $+$0.5 & 0.44 & $-$0.22 & $-$0.31 & \topC{0.18} & 0.14 & $-$5.8 & 18.7 & $-$23 & 64 & $-$0.97 \\
LDA (all articles) & Price + text, LDA (all articles) $\cdot$ \textit{XGBoost} & $+$0.7 & 0.75 & $-$0.95 & $-$0.88 & $-$0.74 & $-$0.60 & $-$12.3 & \topC{13.9} & $-$41 & 122 & $-$1.78 \\
MiniLM & Price + text, MiniLM $\cdot$ \textit{EBM} & $-$0.2 & $-$0.18 & $-$0.13 & $-$0.13 & \topB{0.40} & \topC{0.20} & $-$2.5 & 19.1 & $-$24 & 52 & $-$0.23 \\
FinBERT & Price + text, FinBERT $\cdot$ \textit{EBM} & $+$0.5 & 0.45 & $-$0.00 & 0.20 & \topA{0.69} & \topA{0.99} & 4.1 & 21.1 & $-$16 & 69 & 0.74 \\
NoLBERT-20 & Price + text, NoLBERT-20 $\cdot$ \textit{IPCA} & \topA{$+$3.2} & \topA{2.53} & \topA{0.44} & \topA{0.59} & 0.01 & 0.17 & \topA{10.5} & 17.8 & \topC{$-$14} & 102 & \topC{0.80} \\
\addlinespace[2pt]\midrule
\multicolumn{13}{l}{\textit{Equal-weight combinations}}\\
Price ensemble & Price $\cdot$ \textit{Ensemble} & $+$1.9 & 1.53 & $-$0.04 & $-$0.08 & 0.06 & $-$0.18 & $-$1.6 & 20.9 & $-$27 & 107 & $-$0.40 \\
All readers, 5 d & Price + text, all readers, 5 d $\cdot$ \textit{Ensemble} & $+$1.3 & 1.40 & 0.01 & $-$0.09 & $-$0.21 & $-$0.22 & $-$1.3 & 14.3 & $-$16 & 82 & $-$0.34 \\
All readers, 63 d & Price + text, all readers, 63 d $\cdot$ \textit{Ensemble} & $+$2.0 & 2.03 & 0.30 & 0.39 & $-$0.32 & 0.15 & 5.7 & 14.7 & \topB{$-$13} & 73 & 0.58 \\
All readers, text, 5 d & Text, all readers, 5 d $\cdot$ \textit{Ensemble} & $-$0.8 & $-$0.89 & $-$0.45 & $-$0.57 & $-$0.50 & $-$0.53 & $-$7.1 & \topB{12.4} & $-$21 & \topB{28} & $-$0.94 \\
All readers, text, 63 d & Text, all readers, 63 d $\cdot$ \textit{Ensemble} & $+$0.9 & 1.21 & 0.24 & \topC{0.50} & 0.13 & \topB{0.30} & 5.6 & \topA{11.2} & \topA{$-$10} & \topA{24} & \topA{0.97} \\
Best ten & Ten experiments $\cdot$ \textit{Best ten} & $-$0.4 & $-$0.36 & $-$0.60 & $-$0.43 & $-$0.43 & $-$0.67 & $-$6.5 & 15.2 & $-$23 & 45 & $-$0.66 \\
Price ens. + tone & Price + tone $\cdot$ \textit{Combo} & $+$1.5 & 1.53 & $-$0.72 & $-$0.67 & $-$0.86 & $-$0.70 & $-$9.9 & 14.6 & $-$30 & 128 & $-$2.07 \\
 
\bottomrule
\end{tabular}
\maintablenotes
\end{table}

\begin{table}[!p]
\centering
\scriptsize
\caption{Significance of single portfolios with fixed forecasts, with all benchmarks. SR: Sharpe ratio; $t$, $p$: Newey--West test of the mean return; PSR and DSR: probabilistic and deflated Sharpe ratios. Against a benchmark: Sharpe ratio difference, Ledoit--Wolf test ($t$, $p$) and paired stationary bootstrap ($p$).}
\label{tab:significance-full}
\setlength{\tabcolsep}{3.2pt}
\renewcommand{\arraystretch}{1.06}
\begin{tabular}{@{}L{4.4cm}rrrrr L{2.5cm}rrrr@{}}
\toprule
 & \multicolumn{5}{c}{Against zero} & \multicolumn{5}{c}{Against a benchmark} \\
\cmidrule(lr){2-6}\cmidrule(l){7-11}
Portfolio & SR & $t$ & $p$ & PSR & DSR & Benchmark & $\Delta$SR & LW $t$ & LW $p$ & Boot.\ $p$ \\
\midrule
\multicolumn{11}{l}{\textit{Monthly}}\\
All readers, price + text, 63 d, decile & \topB{1.12} & \topB{2.36} & \topB{0.045} & \topB{0.95} & \topB{0.04} & price ensemble & $+$0.46 & 0.82 & 0.463 & 0.437 \\
 &  &  &  &  &  & same, 5-day shocks & $+$0.74 & 1.64 & 0.149 & 0.155 \\
 &  &  &  &  &  & text alone & $+$0.26 & 0.55 & 0.626 & 0.632 \\
 &  &  &  &  &  & overall pick & \topB{$+$1.35} & 1.45 & 0.196 & 0.159 \\
All readers, price + text, 63 d, buffered & \topA{1.19} & \topA{2.42} & \topA{0.040} & \topA{0.96} & \topA{0.06} & price ensemble & $+$0.66 & 1.24 & 0.267 & 0.237 \\
 &  &  &  &  &  & same, 5-day shocks & $+$0.85 & \topC{1.78} & \topC{0.119} & \topC{0.121} \\
 &  &  &  &  &  & overall pick & \topA{$+$1.56} & 1.65 & 0.146 & 0.110 \\
All readers, text, 63 d, decile & \topC{0.86} & 1.60 & 0.157 & \topC{0.92} & 0.01 & price ensemble & $+$0.20 & 0.23 & 0.836 & 0.842 \\
 &  &  &  &  &  & same, 5-day shocks & $+$1.31 & \topB{2.62} & \topB{0.028} & \topB{0.026} \\
 &  &  &  &  &  & overall pick & $+$1.09 & 1.50 & 0.182 & 0.147 \\
All readers, text, 63 d, buffered & 0.84 & \topC{1.77} & \topC{0.122} & 0.92 & \topC{0.01} & price ensemble & $+$0.31 & 0.40 & 0.722 & 0.716 \\
 &  &  &  &  &  & same, 5-day shocks & \topC{$+$1.32} & \topA{2.68} & \topA{0.025} & \topA{0.021} \\
 &  &  &  &  &  & overall pick & $+$1.21 & 1.73 & 0.129 & 0.093 \\
Price ensemble, decile & 0.66 & 1.50 & 0.183 & 0.84 & 0.01 & price pick & $+$0.90 & 1.08 & 0.334 & 0.316 \\
 &  &  &  &  &  & overall pick & $+$0.89 & 0.88 & 0.432 & 0.407 \\
Overall pick, decile & $-$0.23 & $-$0.33 & 0.767 & 0.35 & 0.00 & price pick & $+$0.02 & 0.03 & 0.980 & 0.976 \\
Overall pick, buffered & $-$0.37 & $-$0.55 & 0.621 & 0.26 & 0.00 & decile rule & $-$0.14 & $-$2.39 & 0.042 & 0.035 \\
\addlinespace[3pt]
\multicolumn{11}{l}{\textit{Biweekly}}\\
All readers, price + text, 63 d, decile & \topC{0.69} & \topC{1.41} & \topC{0.182} & \topC{0.88} & 0.00 & price ensemble & $+$0.59 & 1.17 & 0.266 & 0.217 \\
 &  &  &  &  &  & same, 5-day shocks & $+$0.46 & 1.52 & 0.152 & 0.139 \\
 &  &  &  &  &  & text alone & $+$0.06 & 0.17 & 0.873 & 0.836 \\
 &  &  &  &  &  & overall pick & $+$0.96 & 1.75 & 0.101 & 0.059 \\
All readers, price + text, 63 d, buffered & \topB{0.77} & \topB{1.58} & \topB{0.136} & \topB{0.91} & \topB{0.01} & price ensemble & $+$0.66 & 1.32 & 0.211 & 0.169 \\
 &  &  &  &  &  & same, 5-day shocks & $+$0.40 & 1.25 & 0.236 & 0.218 \\
 &  &  &  &  &  & overall pick & $+$0.90 & 1.54 & 0.147 & 0.111 \\
All readers, text, 63 d, decile & 0.63 & 1.26 & 0.230 & 0.86 & 0.00 & price ensemble & $+$0.53 & 0.83 & 0.428 & 0.358 \\
 &  &  &  &  &  & same, 5-day shocks & \topB{$+$1.10} & \topC{1.83} & \topC{0.088} & \topC{0.066} \\
 &  &  &  &  &  & overall pick & $+$0.89 & 1.77 & 0.098 & 0.073 \\
All readers, text, 63 d, buffered & \topA{0.86} & \topA{1.84} & \topA{0.085} & \topA{0.93} & \topA{0.01} & price ensemble & $+$0.75 & 1.19 & 0.258 & 0.206 \\
 &  &  &  &  &  & same, 5-day shocks & \topA{$+$1.24} & \topA{2.13} & \topA{0.048} & \topA{0.026} \\
 &  &  &  &  &  & overall pick & \topC{$+$0.99} & \topB{2.00} & \topB{0.063} & \topB{0.047} \\
Price ensemble, decile & 0.10 & 0.21 & 0.844 & 0.57 & 0.00 & price pick & $+$0.06 & 0.10 & 0.921 & 0.912 \\
 &  &  &  &  &  & overall pick & $+$0.37 & 0.52 & 0.624 & 0.566 \\
Overall pick, decile & $-$0.27 & $-$0.48 & 0.650 & 0.32 & 0.00 & price pick & $-$0.31 & $-$0.40 & 0.708 & 0.691 \\
Overall pick, buffered & $-$0.13 & $-$0.23 & 0.828 & 0.41 & 0.00 & decile rule & $+$0.14 & 1.13 & 0.283 & 0.303 \\
\addlinespace[3pt]
\multicolumn{11}{l}{\textit{Weekly}}\\
All readers, price + text, 63 d, decile & \topC{0.30} & \topC{0.56} & \topC{0.591} & \topC{0.70} & 0.00 & price ensemble & $+$0.34 & 0.85 & 0.408 & 0.366 \\
 &  &  &  &  &  & same, 5-day shocks & $+$0.29 & 0.83 & 0.422 & 0.438 \\
 &  &  &  &  &  & text alone & $+$0.06 & 0.10 & 0.924 & 0.916 \\
 &  &  &  &  &  & overall pick & \topB{$+$1.37} & \topB{2.16} & \topB{0.040} & \topB{0.038} \\
All readers, price + text, 63 d, buffered & \topB{0.39} & \topB{0.77} & \topB{0.459} & \topB{0.75} & 0.00 & price ensemble & $+$0.46 & 1.22 & 0.239 & 0.203 \\
 &  &  &  &  &  & same, 5-day shocks & $+$0.48 & 1.36 & 0.188 & 0.201 \\
 &  &  &  &  &  & overall pick & \topC{$+$1.37} & \topA{2.38} & \topA{0.024} & \topA{0.025} \\
All readers, text, 63 d, decile & 0.24 & 0.43 & 0.676 & 0.66 & 0.00 & price ensemble & $+$0.28 & 0.33 & 0.748 & 0.711 \\
 &  &  &  &  &  & same, 5-day shocks & $+$0.68 & 1.05 & 0.311 & 0.272 \\
 &  &  &  &  &  & overall pick & $+$1.31 & 1.48 & 0.153 & 0.133 \\
All readers, text, 63 d, buffered & \topA{0.50} & \topA{0.94} & \topA{0.364} & \topA{0.80} & 0.00 & price ensemble & $+$0.57 & 0.71 & 0.489 & 0.430 \\
 &  &  &  &  &  & same, 5-day shocks & $+$1.07 & 1.62 & 0.119 & 0.085 \\
 &  &  &  &  &  & overall pick & \topA{$+$1.48} & 1.76 & 0.092 & 0.076 \\
Price ensemble, decile & $-$0.04 & $-$0.08 & 0.939 & 0.47 & 0.00 & price pick & $-$0.48 & $-$1.50 & 0.148 & 0.116 \\
 &  &  &  &  &  & overall pick & $+$1.03 & \topC{1.90} & \topC{0.070} & \topC{0.060} \\
Overall pick, decile & $-$1.08 & $-$1.88 & 0.073 & 0.04 & 0.00 & price pick & $-$1.51 & $-$2.16 & 0.040 & 0.030 \\
Overall pick, buffered & $-$0.98 & $-$1.73 & 0.097 & 0.05 & 0.00 & decile rule & $+$0.10 & 1.23 & 0.233 & 0.228 \\
 
\bottomrule
\end{tabular}
\tablenotes{\shadekey\ within each frequency; a $p$-value is shaded with its $t$ statistic. ``All readers'': the combination across readers; ``same, 5-day shocks'': that combination with the 5-day window; ``overall pick'' and ``price pick'': the experiments with the highest validation IC overall and on price; ``decile rule'': the same forecast without the buffer.}
\end{table}

\end{document}